\documentclass[]{jfm}
\usepackage{graphicx}
\usepackage{newtxtext}
\usepackage{newtxmath}
\usepackage{natbib}
\usepackage{hyperref}
\hypersetup{
    colorlinks = true,
    urlcolor   = blue,
    citecolor  = black,
}

\newcommand{\RomanNumeralCaps}[1]
\linenumbers

\title{Effects of kinematic and magnetic boundary conditions on the dynamics of convection-driven plane layer dynamos.}

\author{Souvik Naskar\aff{1}, 
  Anikesh Pal\aff{1}
  \corresp{\email{pala@iitk.ac.in}}}

\affiliation{\aff{1}Department of Mechanical Engineering, Indian Institute of Technology,Kanpur 208016, India}

\begin{document}
\maketitle

\begin{abstract}
Rapidly rotating convection-driven dynamos are investigated under different kinematic and magnetic boundary conditions using direct numerical simulations. At a fixed rotation rate, represented by the Ekman number $E=5\times10^{-7}$, the thermal forcing is varied from $2$ to $20$ times its value at the onset of convection ($\mathcal{R}=Ra/Ra_c=2-20$, where $Ra$ is the Rayleigh number), keeping the fluid properties constant ($Pr=Pr_m=1$, where $Pr$ and $Pr_m$ are the thermal and magnetic Prandtl numbers). The statistical behavior, including the force balance and the heat transport characteristics of the dynamos, depend on the boundary conditions that dictate both boundary layer and the interior dynamics. At a fixed thermal forcing ($\mathcal{R}=3$), the horizontal and vertical velocities are higher with thinner thermal boundary layers in no-slip conditions compared to free-slip conditions at the wall. The structure and strength of the magnetic field produced by the dynamo depend on both velocity and magnetic boundary conditions, especially near the walls. Though the leading order force balance remains geostrophic, Lorentz force dominates inside the thermal boundary layer when no-slip, electrically conducting conditions are imposed at the walls. In this case, the work done by the Lorentz force in the turbulent kinetic energy budget is found to have some components that extract energy from the velocity field to produce magnetic field, while some other components extract energy from the magnetic field to produce turbulent kinetic energy. However, with no-slip, perfectly insulated walls, all the components of the work done by the Lorentz force, perform unidirectional energy transfer to produce magnetic energy from the kinetic energy of the fluid to sustain dynamo action. The enhancement of heat transfer in dynamo convection compared to non-magnetic rotating convection exhibits a peak in the range $\mathcal{R}=3-5$ depending on the boundary conditions. In the absence of Ekman layer for free-slip conditions, dynamo action may alter the heat transport significantly by suppressing the formation of large-scale vortices. However, the highest heat transfer enhancement compared to non-magnetic convection occurs when the top and bottom boundaries are no-slip, electrically conducting walls.  
\end{abstract}


%{\bf MSC Codes }  {\it(Optional)} Please enter your MSC Codes here

\section{Introduction}\label{sec:intro}

%Motivations and introduction to dynamo convection
The geomagnetic field act as a shield to protect us from solar wind \citep{tarduno_2018} apart from directly influencing the atmosphere \citep{cnossen_2014}, biology and evolution of life on Earth \citep{erdmann_2021}. Such magnetic fields of planets and stars are known to be generated by a dynamo mechanism driven by convection of electrically conducting fluids \citep{rudiger_2006}. In this self-sustained dynamo mechanism, the convective motion of electrically conducting fluids leads to the amplification of a small magnetic perturbation by electromagnetic induction. The induced magnetic field is then maintained against Joule dissipation by continuously converting some of the kinetic energy of the fluid to magnetic energy. A simple model of a convection-driven dynamo is the Rayleigh-B\'{e}nard convection in a plane layer between two parallel plates, heated from the bottom and cooled from the top, permeated by a magnetic field. Inclusion of global rotation in such flows can stabilize the system by the action of Coriolis force, delaying the onset of convection towards higher thermal forcing \citep{chandrasekhar_1961}. Rotation can also break the reflectional symmetry of dynamo convection to induce large-scale magnetic fields \citep{tobias_2021}. Rotating convection (without magnetic field, henceforth abbreviated as RC) has been studied extensively to investigate their global balance, transport properties, and flow structures using experiments \citep{king_2009,stellmach_2014,kunnen_2010,ecke_2014,king_2013,aurnou_2018,cheng_2020}, direct numerical simulations \citep{guzman_2021,guervilly_2017,kunnen_2016,schmitz_2010,weiss_2010,stellmach_2014,cheng_2015}, and reduced-order asymptotic models \citep{julien_2012a,julien_2012b,rubio_2014,nieves_2014,julien_2016,plumley_2016,plumley_2017,maffei_2021}. Dynamical balances and heat transport in rotating dynamo convection (DC), however, have received less attention.\\

%Introductory remarks on Rotating RBC and the effect on boundary conditions
The flow and thermal field characteristics in a plane layer RC serve as a classical framework for studying solar and planetary convection apart from deep convection in terrestrial oceans \citep{julien_1996}. Here, the convection depends primarily on the thermal forcing, rotation rate, and fluid properties represented by the Rayleigh Number ($Ra$), Ekman Number ($E$), and Prandtl number ($Pr$) respectively (defined in section \ref{sec:method}). The convection begins with steady cellular patterns when the thermal forcing exceeds a critical Rayleigh Number ($Ra_c$), which scales as $E^{-4/3}$ for $Pr>0.67$ in the limit of large rotation rates, $E\xrightarrow{}0$ \citep{chandrasekhar_1961}. This scaling leads to higher $Ra_c$ compared to non-rotating Rayleigh-B\'{e}nard convection, depicting the stabilizing action of the Coriolis force. Increasing the thermal forcing at a fixed rotation rate gives rise to distinct convection regimes with separate flow phenomenology and scaling of the transport properties. The flow regimes are classified as (i) rotation dominated convection, (ii) rotation affected convection, and (iii) rotation unaffected convection, depending on the relative importance of the Coriolis force in the dynamical balance and heat transfer. Rotation-dominated convection is characterized by a geostrophic balance between Coriolis and pressure forces, whereas inertial effects break this balance for rotation affected convection regime at higher thermal forcing. The dependence on the rotation rate is diminished at even higher forcing in the rotation unaffected regime with heat transfer behavior similar to non-rotating Rayleigh-B\'{e}nard convection (see \citet{kunnen_2021} for details). Even the rotation dominated geostrophic convection regime can be divided into sub-regimes with distinct flow structures such as (in the order of increasing thermal forcing) cells, transient Taylor columns, plumes, and large scale vortices (LSV) in the geostrophic turbulent regime \citep{julien_2012b,nieves_2014,kunnen_2016}. Most of these flow features have been confirmed by laboratory experiments in rotating cylinders \citep{cheng_2015,kunnen_2010}. In the simulations, the flow features may also depend on the boundary conditions imposed on the plates. For example, formation of LSVs in geostrophic turbulence is shifted to higher rotation rates with no-slip boundary conditions as compared to the free-slip boundary conditions at a fixed thermal forcing \citep{guzman_2020}, as no-slip boundaries can suppress the formation of LSVs \citep{stellmach_2014,kunnen_2016}. Though the regime transition with thermal forcing is found to be independent of boundary conditions \citep{kunnen_2016}, the force balance and heat transport behavior in RC depends on the kinematic boundary conditions. \\

%Effect of boundary conditions on heat transfer in rotating convection
The classical Rayleigh-B\'{e}nard convection between parallel plates can be separated into two regions: (i) the boundary layer regions with high thermal and velocity gradients near the plates and (ii) the well-mixed bulk region in the interior. In the absence of rotation, the heat transport is throttled by the presence of boundary layers, with the Nusselt number ($Nu$, a non-dimensional measure of heat transfer defined in equation \ref{eqn:nu}) scaling as $Ra^{1/3}$ with the thermal forcing \citep{plumley_2019,iyer_2020}. The thermal behavior of plane layer RC in the rotation-dominated regime is diametrically opposite, with the bulk rather than the boundary layer constraining the convective heat transport. For large rotation rates ($E \rightarrow 0$) the heat transfer should follow the diffusion free scaling $Nu\approx Ra^{3/2}$ irrespective of the boundary conditions \citep{julien_2012a}. Experimental difficulties of maintaining turbulence at small $E$, and computational challenges pertaining to the spatio-temporal resolution requirement restrict the demonstration of this scaling in a laboratory or direct numerical simulations with no-slip boundaries. Instead, Ekman pumping near the thin boundary layers significantly enhances the heat transport even at low Ekman numbers $E \approx 10^{-8}$ \citep{kunnen_2010,stellmach_2014}. This results in a steeper heat transport scaling $Nu\approx Ra^3$, when no-slip conditions are used rather than free-slip conditions at the boundaries. Reduced-order models with parameterized Ekman pumping corroborate these scaling predictions \citep{stellmach_2014,plumley_2017,plumley_2016}. The presence of no-slip walls, with the associated Ekman pumping effect, can significantly enhance vertical velocities, even in the interior, because of the enhanced momentum flux from the boundary towards the bulk. The viscous and inertial force magnitudes near the walls also increase by one order of magnitude compared to their bulk values near the no-slip boundaries leading to increased ageostrophy \citep{guzman_2021}.  \\    

%Literature review on dynamo convection
Motivated by the boundary layer effects on plane layer RC, we intend to investigate the boundary layer dynamics in DC under different combinations of kinematic and magnetic boundary conditions, and their implication on the force balance and heat transport. For rotating DC, the magnetic Prandtl Number ($Pr_m$) appears as an extra parameter that decides the growth and saturation of the magnetic field \citep{tobias_2012,tobias_2021}. Such plane layer convection of electrically conducting fluids was shown to induce dynamo action in early analytical \citep{childress_1972,soward_1974,fautrelle_1982} and numerical studies \citep{meneguzzi_1989}. Using this plane layer model with no-slip and perfectly conducting boundaries, \citet{stpierre_1993} demonstrated subcritical dynamo action at $E=5\times10^{-6}$ with the magnetic field concentrated near the plates. \citet{thelen_2000} studied the effect of vertical, horizontal, and potential magnetic boundary conditions on dynamo action. These boundary conditions were found to dictate the strength and structure of the magnetic field near the plates, though the bulk behavior was independent of the boundary conditions. \citet{stellmach_2004} used free-slip, electrically conducting boundaries to study rapidly rotating ($E=2\times10^{-4}-10^{-6}$), weakly non-linear DC. These particular boundary conditions facilitate comparison of the dynamo behavior with analytical models \citep{childress_1972,soward_1974}. They reported strongly time-dependent flow and magnetic field behavior with cyclic variation between small and large-scale structures. \citet{tilgner_2012,tilgner_2014} reported a transition between large-scale field generation governed by flow helicity to small-scale field generation driven by field stretching. The transition happens at $Re_mE^{1/3}=13.5$ (where $Re_m$ is the magnetic Reynolds number signifying the relative strength of electromagnetic induction over Ohmic diffusion), for electrically conducting boundaries irrespective of the kinematic condition (no-slip or free-slip). Large-scale vortex driven dynamos were demonstrated by \citet{guervilly_2015,guervilly_2017}, that generated large-scale magnetic field. In the absence of a magnetic field, these vortices lead to the reduction of heat transfer between the plates \citep{guervilly_2014}. However small-scale magnetic field may suppress the formation of LSVs at sufficiently high $Re_m>550$ \citep{guervilly_2017}. Asymptotically reduced DC models (vanishingly small inertia with respect to Coriolis force), with leading order geostrophic balance was studied by \citet{calkins_2015}, revealing four distinct dynamo regimes with separate scaling for the magnetic to kinetic energy density ratios \citep{calkins_2018}. At large Prandtl Number ($Pr\rightarrow\infty$), the momentum equation without inertial term becomes linear in velocity, which can be decomposed into the thermally driven part and magnetically driven part \citep{hughes_2016,cattaneo_2017,hughes_2019}. Utilizing this idea \citet{cattaneo_2017}  classified between weak and strong field dynamos based on the relative contribution from the two parts. Rayleigh-B\'enard convection-driven dynamos have been studied by \citet{yan_2021}, who reported heat transfer scaling similar to non-rotating convection. However, the force balance and heat transfer behavior of rotating convection-driven dynamos remain open for exploration.\\%Present objectives and layout of the paper

In the present study, we perform direct numerical simulations of convection-driven dynamos, in the rotation-dominated regime, with varying thermal forcing subjected to different boundary conditions. Our simulations of plane layer RC, with no-slip and free-slip kinematic boundary conditions, serve as references to study the dynamo behavior at four combinations of boundary conditions (combinations of no-slip or free-slip as velocity boundary conditions with perfectly conducting or insulated magnetic boundary conditions). The statistical characteristics of the dynamo, along with the existing force balance in the system, are found to depend on the kinematic and magnetic boundary conditions, both in the bulk and in the boundary layer region. Heat transfer behavior was also found to be strongly dependent on the imposed conditions at the plates. The governing equations with the imposed boundary conditions are detailed in section \ref{sec:method}. The statistical behavior of the flow and magnetic field is presented in section \ref{sec:stat}. In sections \ref{sec:forces} and \ref{sec:budget} we present the force balance and energy budget in the dynamos. Finally, we look into the heat transport behavior in section \ref{sec:heattransfer} and summarize our findings in section \ref{sec:conclusion}.\\

\section{Method}\label{sec:method}

\subsection{Governing Equations}
Dynamo action driven by Rayleigh-B\'{e}nard convection in a three-dimensional Cartesian layer of incompressible, electrically conducting, Boussinesq fluid is considered here. The horizontal layer between two parallel plates has a distance $d$ and temperature difference $\Delta T$, where the lower plate is hotter than the upper plate. The system rotates with an constant angular velocity $\boldsymbol{\Omega}=\Omega\hat{e}_{3}$ about the vertical axis, anti-parallel to the gravity $\boldsymbol{g}=-g\hat{e}_{3}$. The electrically conducting fluid has density $\rho$, kinematic viscosity $\nu$, thermal diffusivity $\kappa$, magnetic permeability $\mu$, electrical conductivity $\sigma$ and the magnetic diffusivity($\lambda$). The layer depth $d$ and temperature difference $\Delta T$ is the natural choice for length and temperature scales, whereas, $u_f=\sqrt{g\alpha\Delta T d}$, and $\sqrt{\rho\mu}u_f$ are chosen to be the velocity \citep{iyer_2020} and magnetic field scales. The non-dimensional governing equations for the velocity field $\boldsymbol{u}$, temperature field $\theta$, and magnetic field $\boldsymbol{B}$ are expressed as the following.
\begin{equation}\label{eqn:solenoidal_nd}
    \frac{\partial u_j}{\partial x_j}=
    \frac{\partial B_j}{\partial x_j}=0,
\end{equation}
\begin{equation}\label{eqn:momentum_nd}
\begin{split}
    \frac{\partial u_i}{\partial t}+
    u_j\frac{\partial u_i}{\partial x_j}=
    -\frac{\partial p}{\partial x_i}+
    \frac{1}{E}\sqrt{\frac{Pr}{Ra}}\epsilon_{ij3} u_j\hat{e_3}+
    B_j\frac{\partial B_i}{\partial x_j}+
    \theta\delta_{i3}+
    \sqrt{\frac{Pr}{Ra}}\frac{\partial^{2} u_{i}}{\partial x_j\partial x_j},
\end{split}
\end{equation}
\begin{equation}\label{eqn:energy_nd}
    \frac{\partial \theta}{\partial t}+
    u_j\frac{\partial \theta}{\partial x_j}=
    \frac{1}{\sqrt{RaPr}}\frac{\partial^2\theta}{\partial x_j\partial x_j},
\end{equation}
\begin{equation}\label{eqn:maxwell_nd}
    \frac{\partial B_i}{\partial t}+
    u_j\frac{\partial B_i}{\partial x_j}=
    B_j\frac{\partial u_i}{\partial x_j}+
    \sqrt{\frac{Pr}{Ra}}\frac{1}{Pr_m}\frac{\partial^{2} B_{i}}{\partial x_j\partial x_j}.
\end{equation}

The definitions of the four non-dimensional parameters, namely Rayleigh Number ($Ra$), Ekman Number ($E$), kinematic and magnetic Prandtl numbers ($Pr$ and $Pr_m$) are given as follows.
\begin{equation}\label{eqn:nd_parameters}
    Ra=\frac{g\alpha\Delta T d^3}{\kappa\nu},\quad E=\frac{\nu}{2\Omega d^2},\quad Pr=\frac{\nu}{\kappa},\quad Pr_m=\frac{\nu}{\lambda}. 
\end{equation}

 In the horizontal directions ($x_{1},x_{2}$) periodic boundary conditions are applied. As we aim to study the effect of boundary layer dynamics on the dynamo convection, both no-slip and free-slip boundary conditions are implemented in the vertical direction ($x_3$) as follows:

\begin{equation}\label{eqn:ubc}
\begin{split}
    u_{1}=u_{2}=u_{3}=0 \; \textrm{ at } \; x_{3}=\pm1/2 \qquad\text{(no-slip)} \\
   \frac{\partial u_{1}}{\partial x_{3}}=\frac{\partial  u_{2}}{\partial x_{3}}=0,\  u_{3}=0 \  \textrm{ at } \  x_{3}=\pm1/2 \qquad\text{(free-slip)}.    
\end{split}
\end{equation} 

Isothermal boundary conditions with unstable temperature gradient are imposed to drive convection as follows.

\begin{equation}\label{eqn:tbc}
\begin{split}
    \quad \theta=1/2 \; \textrm{ at } \; x_{3}=-1/2, \quad \theta=-1/2 \; \textrm{ at } \; x_{3}=1/2
\end{split}
\end{equation}

For the magnetic field, both perfectly conducting and perfectly insulating boundary conditions are implemented to compare the resulting magnetic field structure. For the insulated boundary we have imposed all the magnetic field components to be zero at the boundaries whereas, for conducting boundary the field is constrained to be horizontal at the wall \citep{jones_2000,cattaneo_2006}.

\begin{equation}\label{eqn:mbc}
\begin{split}
    B_{1}=B_{2}=B_{3}=0 \; \textrm{ at } \; x_{3}=\pm1/2 \qquad\text{(insulating)} \\
   \frac{\partial B_{1}}{\partial x_{3}}=\frac{\partial  B_{2}}{\partial x_{3}}=0,\  B_{3}=0 \  \textrm{ at } \  x_{3}=\pm1/2 \qquad\text{(conducting)}   
\end{split}
\end{equation} 

\subsection{Simulation Details}\label{sec:simulation}

\begin{table}
  \begin{center}
\def~{\hphantom{0}}
  \begin{tabular}{lccccccccccccc}
       $\mathcal{R}$  &  $\widetilde{Ra}$   &  $Ro_C$ &  $\Lambda_{tr}$ &  $\Lambda_V$ &  $\Lambda_T$ &  $Re_m$ &  $\frac{Nu}{Nu_0}$ &  $\frac{\langle\epsilon_v\rangle}{\langle\epsilon_0\rangle}$ &  $\frac{\langle\epsilon_j\rangle}{\langle\epsilon\rangle}$ &  $\langle\mathcal{B}\rangle^{*}$ &  $-\langle\epsilon\rangle^{*}$ &  $\tau$ &  $\tau_{0}$ \\[3pt]
       2   & 15.2 & 0.031 & 0.219 & 0.038 & 0.005 & 729 & 0.73 & 0.50 & 0.32 & 0.562 & 0.542 & 1.171 & -\\
       2.5   & 19.0 & 0.035 & 0.516 & 0.011 & 0.292 & 1327 & 1.26 & 0.94 & 0.34 & 1.955 & 1.934 & 1.027 & -\\
       3   & 22.8 & 0.038 & 1.031 & 0.071 & 0.617 & 2012 & 1.72 & 1.44 & 0.36 & 3.566 & 3.459 & 0.933 & -\\
       4   & 30.4 & 0.044 & 1.683 & 0.087 & 0.555 & 2284 & 1.49 & 1.23 & 0.38 & 3.737 & 3.694 & 0.848 & -\\
       5   & 38.0 & 0.049 & 2.399 & 0.129 & 0.462 & 2660 & 1.43 & 1.08 & 0.39 & 3.675 & 3.661 & 0.827 & -\\
       10   & 76.0 & 0.069 & 3.743 & 0.132 & 0.214 & 4203 & 1.17 & 0.85 & 0.34 & 3.902 & 3.818 & 0.784 & -\\
       20   & 152.0 & 0.098 & 13.287 & 0.183 & 0.167 & 7642 & 1.14 & 0.82 & 0.39 & 5.141 & 4.888 & 0.930 & -\\       
  \end{tabular}
  \caption{Statistics of the dynamo simulations with NSC boundary conditions.} 
  \label{tab:nsc}
  \hfill\break
%  \hfill\break]
  \begin{tabular}{lcccccccccccccc}
       $\mathcal{R}$  &  $\widetilde{Ra}$   &  $Ro_C$ &  $\Lambda_{tr}$ &  $\Lambda_V$ &  $\Lambda_T$ &  $Re_m$ &  $\frac{Nu}{Nu_0}$ &  $\frac{\langle\epsilon_v\rangle}{\langle\epsilon_0\rangle}$ &  $\frac{\langle\epsilon_j\rangle}{\langle\epsilon\rangle}$ &  $\langle\mathcal{B}\rangle^{*}$ &  $-\langle\epsilon\rangle^{*}$ &  $\tau$ &  $\tau_{0}$ \\[3pt]
       2   & 15.2 & 0.031 & 0.588 & 0.001 & 0.017 & 523 & 0.67 & 0.47 & 0.38 & 0.486 & 0.468 & 1.323 & 1.058\\
       2.5   & 19.0 & 0.035 & 0.015 & 0.055 & 0.056 & 1350 & 1.02 & 1.24 & 0.21 & 1.562 & 1.545 & 1.053 & 1.071\\
       3   & 22.8 & 0.038 & 0.268 & 0.101 & 0.109 & 1822 & 1.19 & 1.26 & 0.13 & 2.421 & 2.421 & 1.015 & 1.046\\
       4   & 30.4 & 0.044 & 0.660 & 0.115 & 0.151 & 2305 & 1.27 & 1.17 & 0.24 & 3.166 & 3.130 & 0.903 & 0.986\\
       5   & 38.0 & 0.049 & 0.614 & 0.058 & 0.152 & 2570 & 1.21 & 1.08 & 0.23 & 3.092 & 3.080 & 0.847 & 0.958\\
       10   & 76.0 & 0.069 & 0.821 & 0.033 & 0.193 & 3952 & 1.07 & 0.87 & 0.29 & 3.571 & 3.494 & 0.770 & $1.102^{\#}$\\
       20   & 152.0 & 0.098 & 1.422 & 0.021 & 0.171 & 6783 & 1.06 & 0.80 & 0.37 & 4.810 & 4.665 & 0.775 & $1.341^{\#}$\\       
  \end{tabular}
  \caption{Statistics of the dynamo simulations with NSI boundary conditions.} 
  \label{tab:nsi}
  \hfill\break
%  \hfill\break  
  \begin{tabular}{lccccccccccccc}
       $\mathcal{R}$  &  $\widetilde{Ra}$   &  $Ro_C$ &  $\Lambda_{tr}$ &  $\Lambda_V$ &  $\Lambda_T$ &  $Re_m$ &  $\frac{Nu}{Nu_0}$ &  $\frac{\langle\epsilon_v\rangle}{\langle\epsilon_0\rangle}$ &  $\frac{\langle\epsilon_j\rangle}{\langle\epsilon\rangle}$ &  $\langle\mathcal{B}\rangle^{*}$ &  $-\langle\epsilon\rangle^{*}$ &  $\tau$  &  $\tau_{0}$ \\[3pt]
       2   & 17.4 & 0.033 & 0.034 & 0.001 & 0.001 & 641 & 0.98 & 0.98 & 0.07 & 0.345 & 0.343 & 1.174 & 1.351\\
       2.5   & 21.7 & 0.037 & 0.017 & 0.002 & 0.002 & 950 & 1.00 & 0.97 & 0.03 & 0.524 & 0.520 & 1.234 & 1.350\\
       3   & 26.1 & 0.041 & 0.017 & 0.003 & 0.003 & 1176 & 1.08 & 1.02 & 0.01 & 0.585 & 0.581 & 1.171 & 1.414\\
       4   & 34.8 & 0.047 & 1.005 & 0.004 & 0.006 & 2087 & 1.44 & 1.04 & 0.34 & 1.812 & 1.753 & 1.019 & $2.960^{\#}$\\
       5   & 43.5 & 0.052 & 1.488 & 0.005 & 0.013 & 2562 & 1.42 & 1.02 & 0.35 & 2.248 & 2.162 & 0.953 & $3.054^{\#}$\\
       10   & 87.0 & 0.074 & 4.145 & 0.007 & 0.037 & 3957 & 1.02 & 0.74 & 0.39 & 2.883 & 2.718 & 0.825 & $7.755^{\#}$\\
       20   & 174.0 & 0.105 & 13.140 & 0.009 & 0.091 & 7249 & 0.98 & 0.73 & 0.42 & 4.482 & 4.261 & 0.802 & $4.050^{\#}$\\       
  \end{tabular}
  \caption{Statistics of the dynamo simulations with FSC boundary conditions.} 
  \label{tab:fsc}
  \hfill\break
%  \hfill\break  
  \begin{tabular}{lccccccccccccc}
       $\mathcal{R}$  &  $\widetilde{Ra}$   &  $Ro_C$ &  $\Lambda_{tr}$ &  $\Lambda_V$ &  $\Lambda_T$ &  $Re_m$ &  $\frac{Nu}{Nu_0}$ &  $\frac{\langle\epsilon_v\rangle}{\langle\epsilon_0\rangle}$ &  $\frac{\langle\epsilon_j\rangle}{\langle\epsilon\rangle}$ &  $\langle\mathcal{B}\rangle^{*}$ &  $-\langle\epsilon\rangle^{*}$ &  $\tau$  &  $\tau_{0}$ \\[3pt]
       2   & 17.4 & 0.033 & 0.001 & 0.009 & 0.007 & 664 & 0.96 & 1.04 & 0.01 & 0.338 & 0.337 & 1.248 & -\\
       2.5   & 21.7 & 0.037 & 0.002 & 0.014 & 0.008 & 979 & 0.97 & 0.99 & 0.02 & 0.524 & 0.520 & 1.319 & -\\
       3   & 26.1 & 0.041 & 0.006 & 0.025 & 0.019 & 1157 & 1.07 & 1.00 & 0.08 & 0.578 & 0.574 & 1.288 & -\\
       4   & 34.8 & 0.047 & 0.968 &  0.061 & 0.025 & 2117 & 1.20 & 0.88 & 0.31 & 1.496 & 1.473 & 1.145 & -\\
       5   & 43.5 & 0.052 & 1.415 & 0.012 & 0.053 & 2469 & 1.21 & 0.86 & 0.34 & 1.905 & 1.851 & 1.039 & -\\
       10   & 87.0 & 0.074 & 3.885 & 0.026 & 0.066 & 3978 & 0.95 & 0.70 & 0.37 & 2.686 & 2.525 & 0.870 & -\\
       20   & 174.0 & 0.105 & 11.649 & 0.053 & 0.121 & 7587 & 0.71 & 0.71 & 0.40 & 4.420 & 4.397 & 0.859 & -\\       
  \end{tabular}
  \caption{Statistics of the dynamo simulations with FSI boundary conditions. In these tables, $\widetilde{Ra}=RaE^{4/3}$ is the reduced Rayleigh number and $Ro_C=E\sqrt{Ra/Pr}$ is the convective Rossby Number. $\Lambda_{tr}=\sigma B_\tau^2/\rho\Omega$ is the traditional definition of Elsasser number, where $B_\tau$ is the magnetic field scale. $\Lambda_V$ is the volume-averaged Elsasser number. $\Lambda_T$ is the Elsasser number at the thermal boundary layer edge.$Re_m=u_\tau d/\lambda$, is the magnetic Reynolds number, where $u_\tau=\langle 2K \rangle^{1/2}$ is the velocity scale. $Nu/Nu_0$ and $\langle\epsilon_v\rangle/\langle\epsilon_0\rangle$ are the Nusselt number ratio and viscous dissipation ratio. $\langle\epsilon_j\rangle/\langle\epsilon\rangle$ is the ratio of Joule dissipation to total dissipation. The buoyancy flux $\langle\mathcal{B}\rangle^{*}=\langle u_3\theta\rangle\times10^4$ and total dissipation $\langle\epsilon\rangle^{*}=\langle\epsilon_{v}+\epsilon_{j}\rangle\times10^{4}$ depicts the overall balance of energy. $\tau$ is the kinetic energy ratio indicating presence of LSVs. Non-magnetic simulations with LSVs are indicated by $\#$ in the last column.} 
  \label{tab:fsi}  
  
  \end{center}
\end{table}

The governing equations \ref{eqn:solenoidal_nd}-\ref{eqn:maxwell_nd} are solved in a cubic domain with unit side length, using finite difference method. The geometrical details and numerical algorithms are presented in \citet{naskar_2021}. We investigate the dynamical balances and heat transport in DC at constant rotation rate and constant fluid properties with variation in thermal forcing for different boundary conditions. The thermal forcing is represented by the convective supercriticality $\mathcal{R}=Ra/Ra_c$ where $Ra_c$ is the minimum required value of $Ra$ to start steady rotating convection \citep{chandrasekhar_1961}. In this study, we have used the values of critical Rayleigh Number for non-magnetic convection as $Ra_c=8.6E^{-4/3}$ for free-slip \citep{chandrasekhar_1961} and $Ra_c=7.6E^{-4/3}$ for no-slip boundaries \citep{king_2012,kunnen_2021}. We choose the values of $\mathcal{R}=2,2.5,3,4,5,10,20$, Ekman number $E=5\times10^{-7}$ and the Prandtl Numbers $Pr=Pr_m=1$ for the present simulations. To investigate dynamo action for different boundary conditions, we perform six simulations at each value of $\mathcal{R}$: (a) non-magnetic rotating convection with no-slip (NS) and free-slip (FS) boundary conditions, and (b) dynamo simulations at with no-slip and free-slip boundary conditions with perfectly conducting (NSC and FSC) and perfectly insulated boundaries (NSI and FSI). The simulation inputs and diagnostic parameters are summarized in tables \ref{tab:nsc}-\ref{tab:fsi}. A mesh with $1024\times1024\times256$ grid points is used for all the simulations, with uniform spacing in the horizontal and grid clustering in the vertical direction to resolve the boundary layers. The solver has been extensively validated for studies on rotating convection \citep{pal_2020}, and various transitional and turbulent shear flows \citep{Pal2013,Pal2015,brucker_2010,Pham_2009}. Details of the grid resolution and validation studies are reported in a previous study \citep{naskar_2021}. The scaled values of the buoyancy flux, $\langle\mathcal{B}\rangle^{*}=\langle u_3\theta\rangle\times10^4$ and the total dissipation $\langle\epsilon\rangle^{*}=\langle\epsilon_{v}+\epsilon_{j}\rangle\times10^{4}$ in tables \ref{tab:nsc}-\ref{tab:fsi} indicates sufficient resolution for all our simulations, as the grid can capture most of the energetic scales. It should be noted that the combination of non-dimensional numbers appearing before the Coriolis term in the momentum equation is the inverse of convective Rossby Number $Ro_C=E(Ra/Pr)^{1/2}$ frequently used in the literature on rapidly rotating convection \citep{aurnou_2020}. For all our simulations, the Convective Rossby number $Ro_C\ll1$ indicates rapidly rotating convection regime, as shown in tables \ref{tab:nsc}-\ref{tab:fsi}. The reduced Rayleigh number $\widetilde{Ra}=RaE^{4/3}$ is another important parameter presented in these tables to compare against the literature on rapidly rotating convection \citep{julien_2012a,king_2012,calkins_2018}. 

\subsection{Turbulence statistics}
Reynolds decomposition is performed on all the variables. For moderate to high $Re_m\geq O(10-100)$, the system can induce its own magnetic field with a wide range of length and time scales. In such cases, it is worthwhile to decompose the magnetic field into mean and fluctuating parts following the developments in mean-field electrodynamics \citep{cattaneo_2006}.

\begin{equation}\label{eqn:def_average}
\begin{split}
    \Bar{\phi}(z,t)=\int_{A_h}\phi(x,y,z,t)dxdy\\
    \phi(x,y,z,t)=\Bar{\phi}(z,t)+\phi '(x,y,z,t)
\end{split}    
\end{equation}

where $A_h$ is the horizontal area of integration of the flow variables $\phi=\{u_i, p, \theta, B_i\}$. The \textit{r.m.s.} values can also be calculated as $\phi_{rms}=\left(\overline{\phi^{\prime^{2}}}\right)^{1/2}$. The associated energy budgets for the turbulent kinetic energy($K$) is presented in Appendix \ref{app:Budget}. \\

We compare our simulations in terms of the heat transfer, represented by the Nusselt number($Nu$). This is defined as the total heat flux to the conductive heat flux transferred from the bottom plate to the top plate.

\begin{equation}\label{eqn:nu}
\begin{split}
    Nu=\frac{qd}{k\Delta T}=1+\sqrt{RaPr}\langle\mathcal{B}\rangle
\end{split}
\end{equation}

Here $\langle\phi\rangle=\int_{0}^{1}\overline{\phi}\,dx_{3}$ denote average over the entire the volume. The volume-averaged total heat flux and vertical buoyancy flux of energy are denoted by $q$ and $\langle\mathcal{B}\rangle$ respectively. Subscript $"0"$ is used to represent the properties without magnetic field (NS and FS cases) in the rest of this paper. All statistics presented here are averaged in time for more than 100 free-fall time units, after the simulations settle in a statistically stationary state.  

\section{Result}\label{sec:results}

\citet{naskar_2021} performed DNS of rapidly-rotating dynamos with no-slip boundary conditions and reported a significant enhancement ($72\%$) in heat transfer as compared to non-magnetic rotating convection at $\mathcal{R}=3$. An increase in the Lorentz force near the boundaries was found to be the reason for this enhanced heat transport. Owing to this interesting behavior, we study the statistical details, force balance and energy budget of the dynamos at $\mathcal{R}=3$ subjected to different boundary conditions. To further understand the changes in the dynamo behavior with $\mathcal{R}$, we have tabulated the volume-averaged statistics in tables \ref{tab:nsc}-\ref{tab:fsi}. The heat transfer behavior for all our simulations are summarized in section \ref{sec:heattransfer}.

\subsection{Statistical details of the dynamos}\label{sec:stat}

\begin{figure*}
\centering
(a) \includegraphics[width=0.482\linewidth,trim={0.2cm 6.2cm 0cm 9cm},clip]{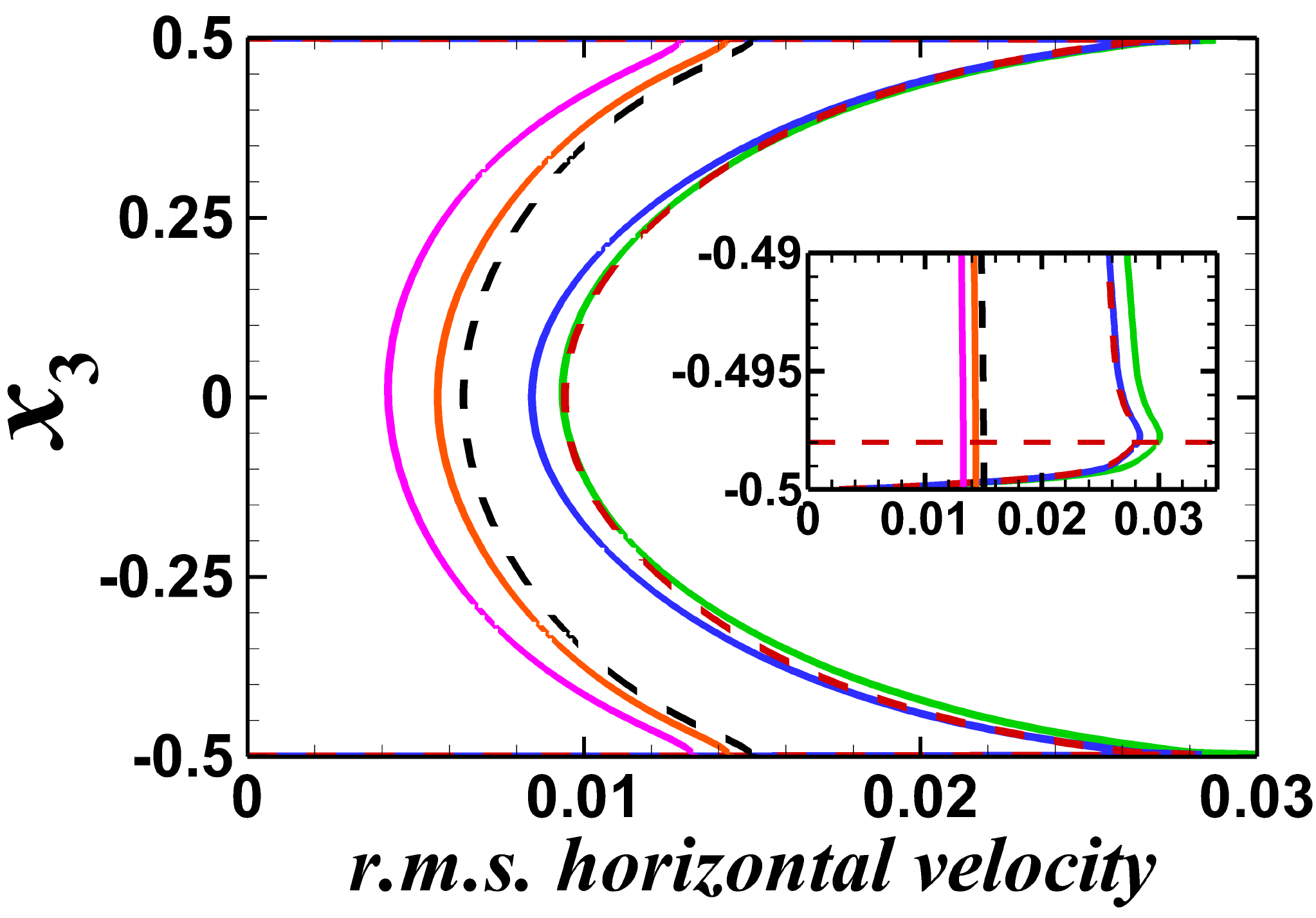}
(b) \includegraphics[width=0.448\linewidth,trim={1.8cm 6.2cm 0cm 9cm},clip]{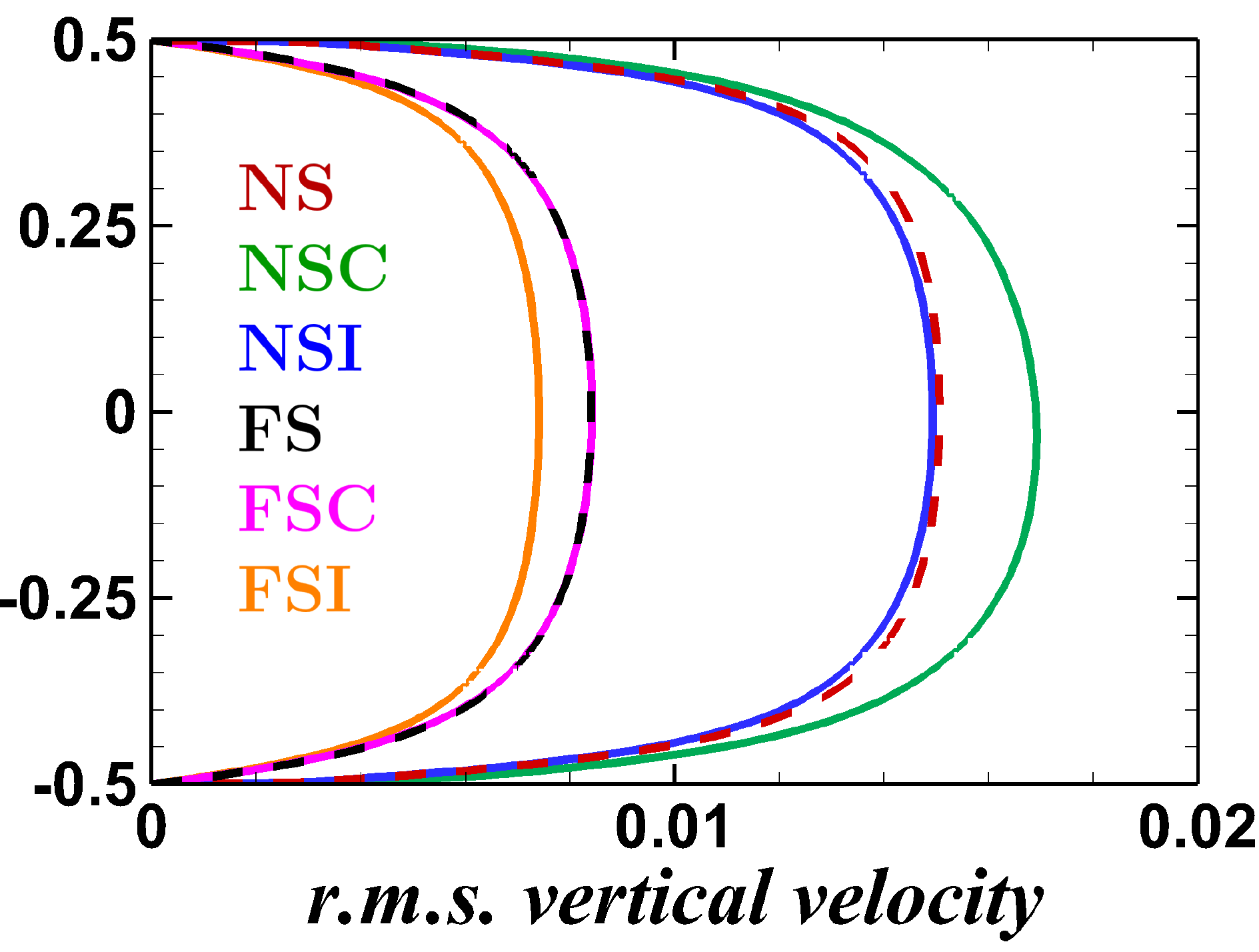}\\
(c) \includegraphics[width=0.482\linewidth,trim={0.2cm 6.2cm 0cm 9cm},clip]{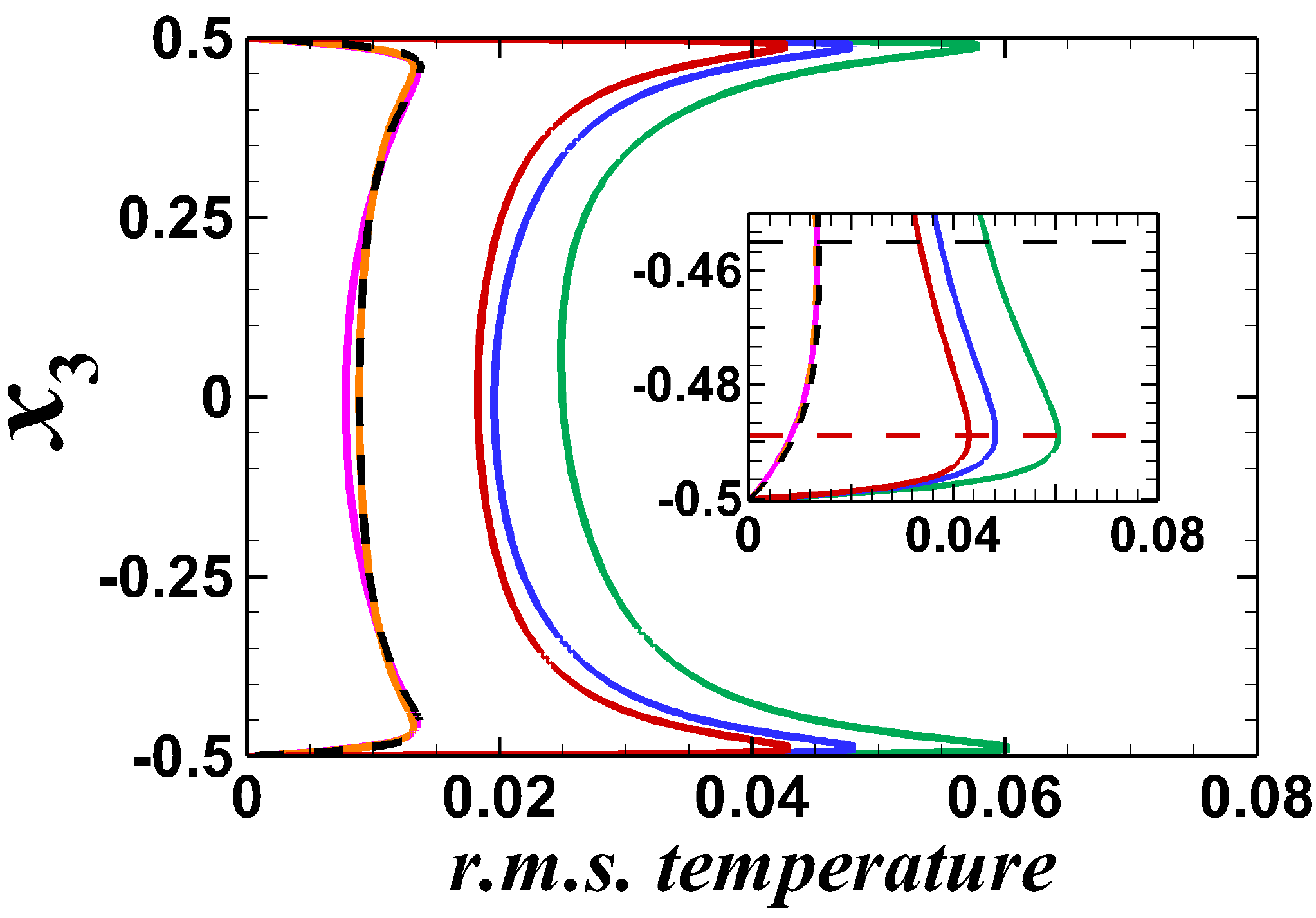}
(d) \includegraphics[width=0.448\linewidth,trim={1.8cm 6.2cm 0cm 9cm},clip]{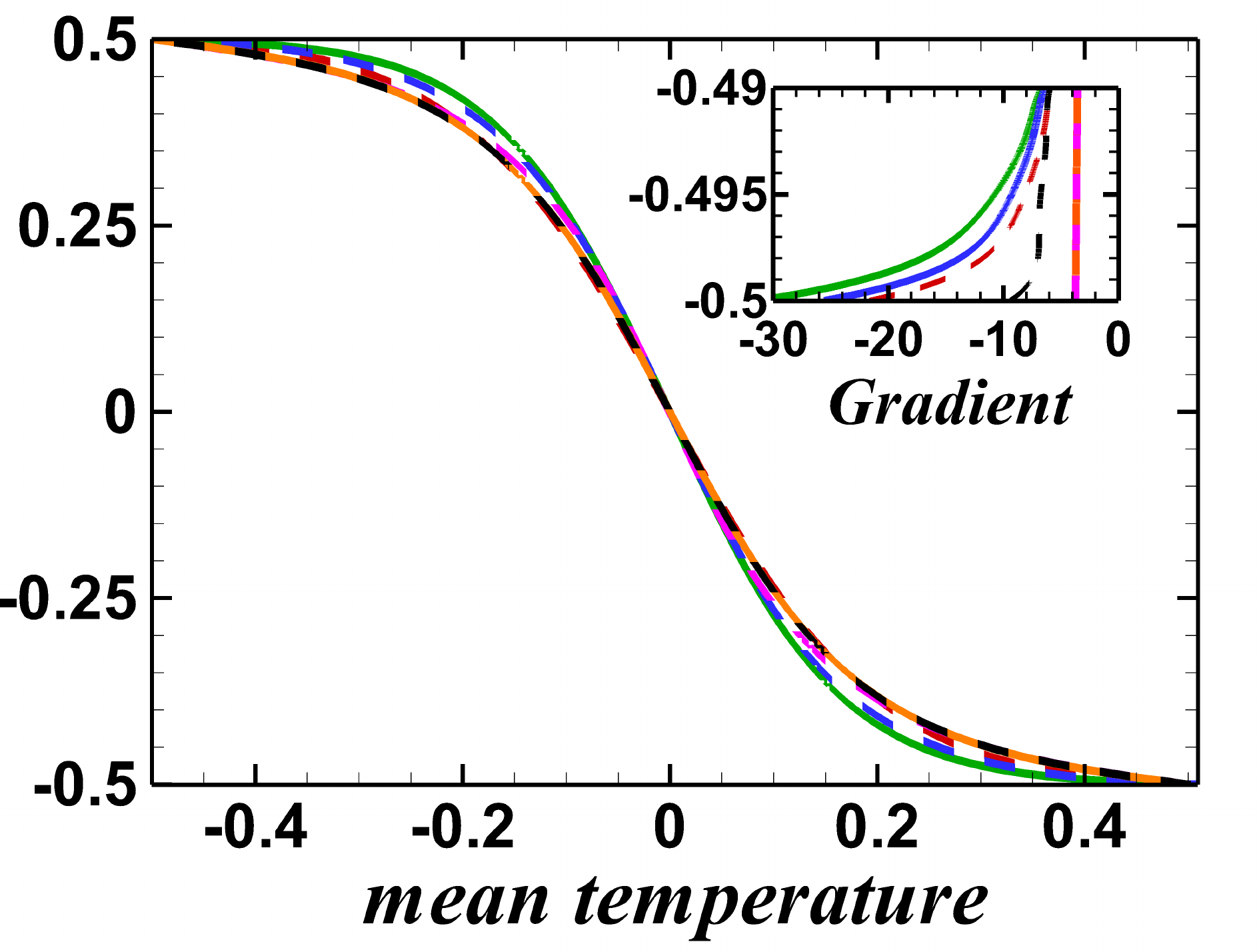}

\caption{Vertical variation of \textit{r.m.s.} quantities: (\textit{a}) Horizontal velocity, (\textit{b}) Vertical velocity, (\textit{c}) \textit{r.m.s.} temperature, (\textit{d})  mean temperature at $\mathcal{R}=3$. All quantities are averaged in time and in the horizontal directions. Dashed lines are used as needed for improving clarity of the plots.}
\label{fig:vel}
\end{figure*}

\begin{figure}%[tbhp]
\centering
(a) \includegraphics[width=0.94\linewidth,trim={0cm 8cm 0cm 11cm},clip]{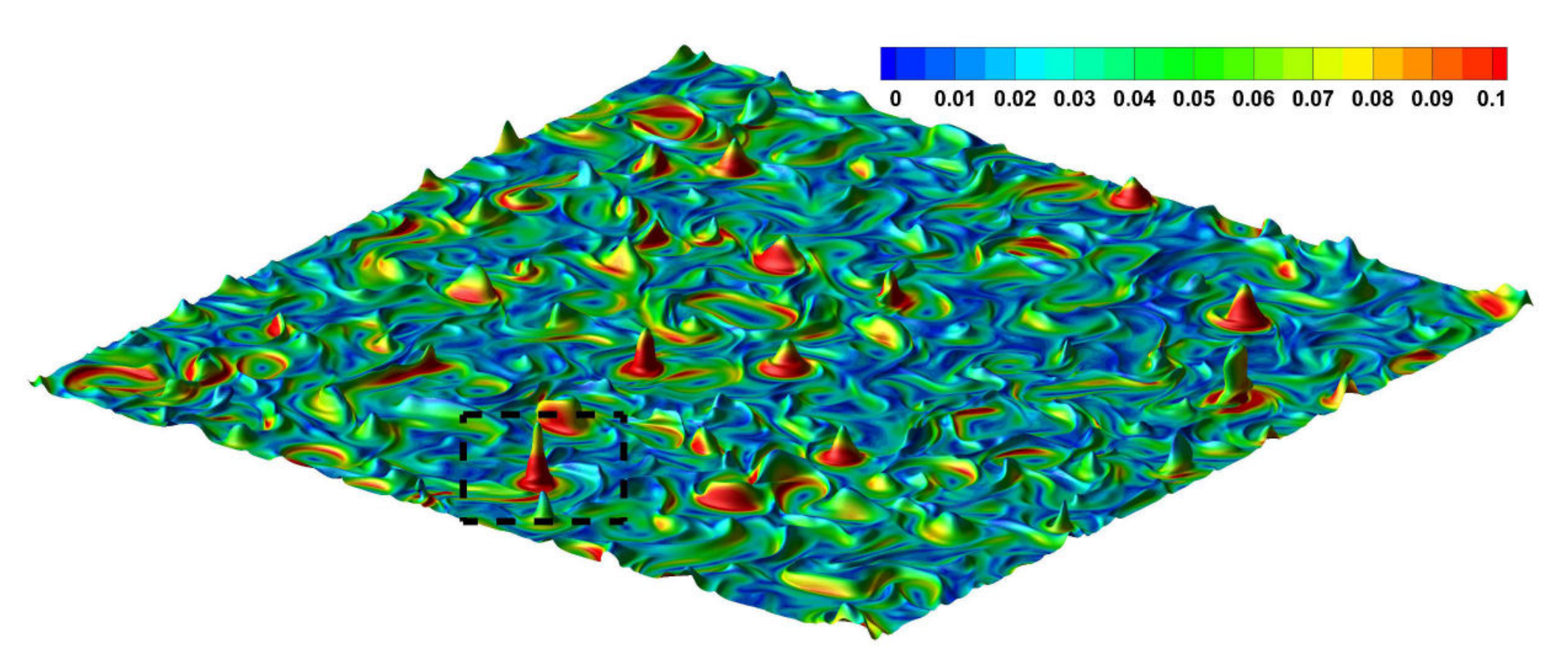}\\
(b) \includegraphics[width=0.465\linewidth,trim={0.2cm 6cm 0cm 6cm},clip]{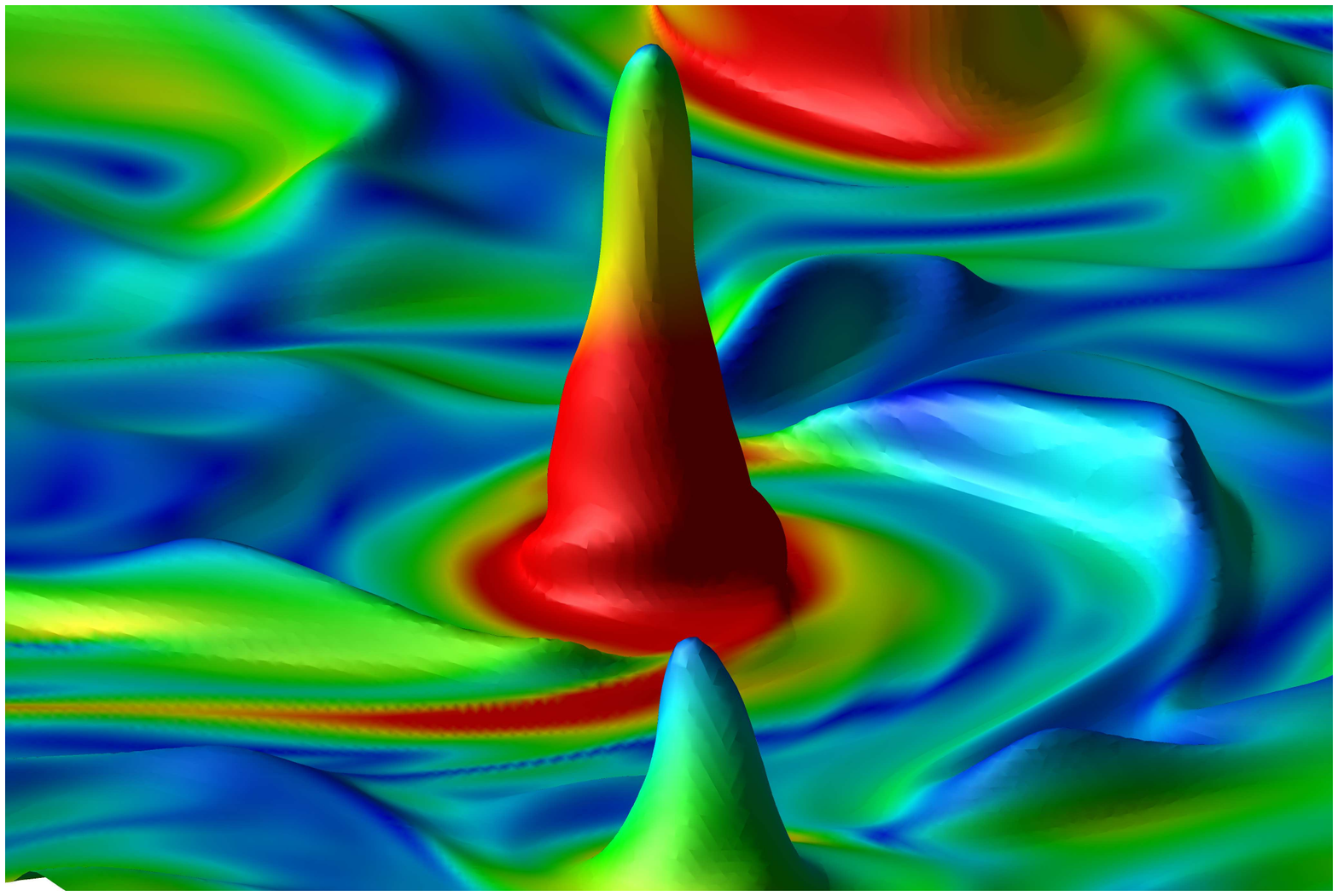}
(c) \includegraphics[width=0.465\linewidth,trim={0.2cm 6cm 0cm 6cm},clip]{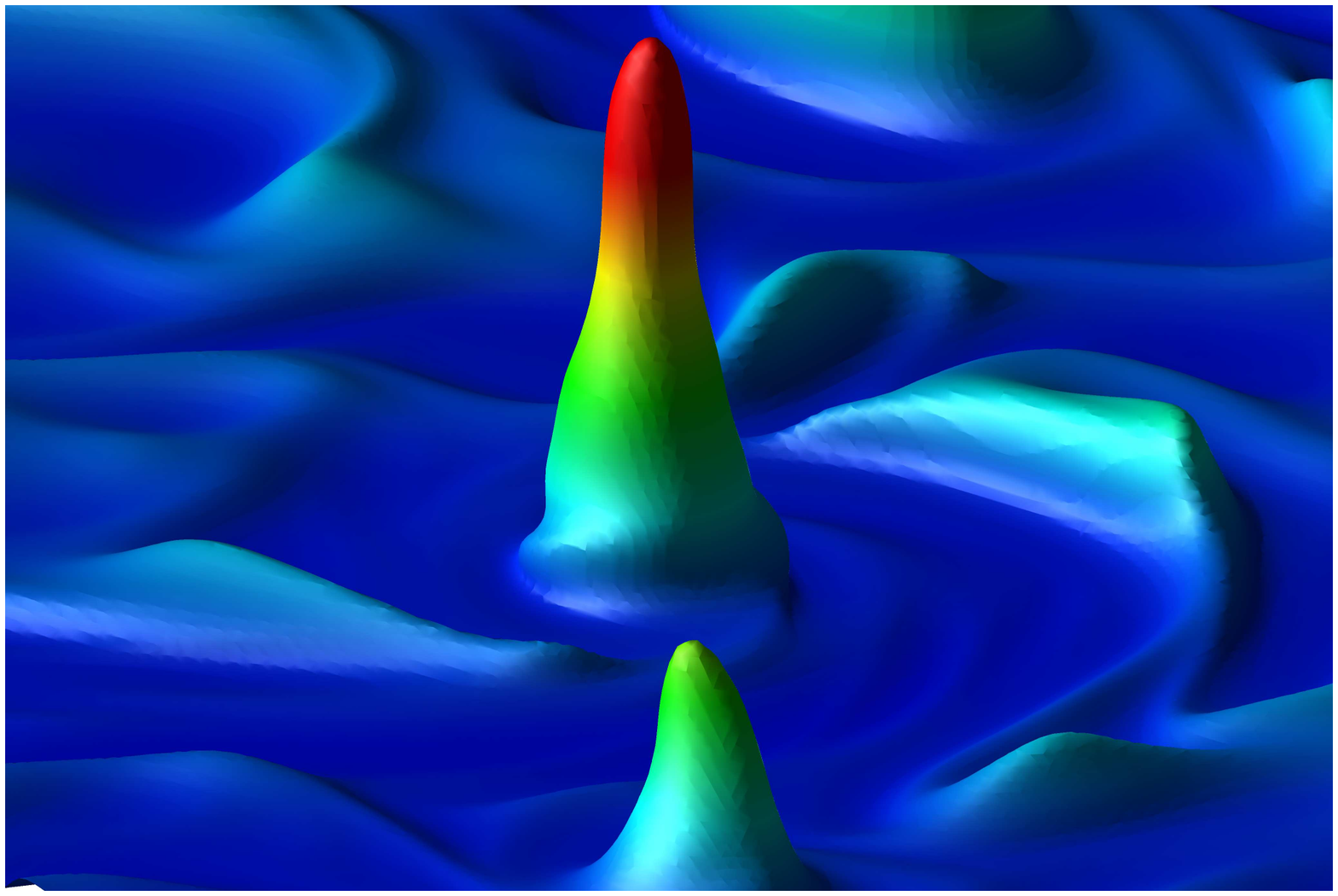}
\caption{Plume sites visualized by isosurface of instantaneous temperature field with $\theta=0.45$, colored by horizontal velocity for NSC case at $\mathcal{R}=3$. A typical plume site marked by dashed rectangle has been magnified and colored by horizontal and vertical velocities in figures (b) and (c) respectively.}
\label{fig:plume}
\end{figure}

In this section, we discuss the statistical behavior of the velocity, temperature, and magnetic field of the dynamos subjected to different boundary conditions at $\mathcal{R}=3$. In figure \ref{fig:vel}a, the \textit{r.m.s.} horizontal velocity is presented. To clarify the near-wall variation, we have included a magnified inset. At this point, it is important to distinguish between the well-mixed bulk region in the interior and the boundary layer region with high gradients near the plates. Therefore, we define the thermal boundary layer as the region near the plate where temperature gradients are high, and its thickness ($\delta_T$) is evaluated as the distance from the wall where the \textit{r.m.s.} value of temperature reaches a maximum \citep{king_2009}. Furthermore, when the no-slip condition is imposed, the viscous effects are confined within a thin Ekman layer, defined by the distance of the maximum of horizontal \textit{r.m.s.} velocity from the wall, $\delta_E$. The edge of the Ekman boundary layer for the no-slip cases at $x_{3}=-0.498$, as marked with a horizontal red dashed line in the inset in figure \ref{fig:vel}a, remains independent of the magnetic boundary conditions. In figure \ref{fig:vel}a, the horizontal velocity in the bulk for NS and NSC cases overlap, whereas, near the Ekman layer, the velocities for NS and NSI cases show the same behavior. For the no-slip boundary condition, the horizontal velocity can be seen to be higher than that of free-slip boundaries, both in the bulk and near the boundaries. A similar behavior is observed for \textit{r.m.s.} vertical velocity and temperature fluctuations in figure \ref{fig:vel}b and c respectively. The vertical variation of \textit{r.m.s.} velocities can be understood from the Ekman pumping mechanism \citep{guzman_2021}.The velocity magnitudes around a plume site as depicted in figure \ref{fig:plume} can provide further insight into this phenomenon. Here, the conical plume sites can be recognized from the temperature isosurface $\theta=0.45$ near the lower boundary in figure \ref{fig:plume}a. The horizontal convergence (or divergence) of fluid at the sites of the vortical plumes (figure \ref{fig:plume}b) enhances horizontal velocity near the wall with no-slip boundary condition compared to free-slip cases in figure \ref{fig:vel}a. This fluid then gains vertical acceleration towards the bulk, resulting in higher vertical velocities (figure \ref{fig:plume}c), as plotted in figure \ref{fig:vel}b. Ekman pumping induced by the no-slip boundaries is known to enhance momentum and heat transport \citep{stellmach_2014}, and is the reason for enhanced \textit{r.m.s.} velocities and temperature fluctuations, especially near the boundaries. Furthermore, in the inset of figure \ref{fig:vel}c, the thermal boundary layer thickness ($\delta_T$) for free-slip boundaries (horizontal black dashed line) is more than four times higher than no-slip boundaries (horizontal red dashed line). The thermal fluctuations are enhanced with no-slip boundary conditions, with maximum \textit{r.m.s.} fluctuation shifting towards the wall. It is noteworthy that changing the boundary conditions can significantly modulate the bulk behavior apart from the boundary layer dynamics. Also, the effects of changing the kinematic boundary condition on the velocity and thermal fields are more prominent than the magnetic conditions. The mean temperature profile shows a higher temperature gradient near the bottom wall for no-slip conditions (see the inset at the upper right corner of figure \ref{fig:vel}d), with the highest vertical gradient for the NSC case indicating the highest heat transfer from the wall among all the cases at $\mathcal{R}=3$ (see section \ref{sec:heattransfer} for detailed discussion). However, the mean temperature profile and its gradient at the mid-plane remain nearly independent of boundary conditions. The magnetic Reynolds number $Re_m=RePr_m=u_\tau d/\lambda$, (where $u_\tau=\langle 2K \rangle^{1/2}$ is the velocity scale), is same as the Reynolds number $Re$, for $Pr_m=1$ in the present simulations (see tables \ref{tab:nsc}-\ref{tab:fsi}). The Reynolds number increases by an order of magnitude in the range $\mathcal{R}=2-20$, indicating increased velocity fluctuations with thermal forcing, irrespective of boundary conditions. The \textit{r.m.s.} temperature fluctuations also increase monotonically with increased thermal forcing (figure not presented). The decreasing rotational constraint with increasing $\mathcal{R}$ leads to decreasing effect of Ekman pumping on the velocity and temperature field. Therefore, the difference between \textit{r.m.s.} velocity and temperature magnitudes with no-slip and free-slip conditions diminish with increasing $\mathcal{R}$. For the FS cases with LSVs, horizontal velocity becomes larger than any other cases (see tables \ref{tab:nsi}-\ref{tab:fsc} and section \ref{sec:heattransfer} for details). \\

\begin{figure*}
\centering
(a) \includegraphics[width=0.482\linewidth,trim={0.2cm 6.2cm 0cm 9cm},clip]{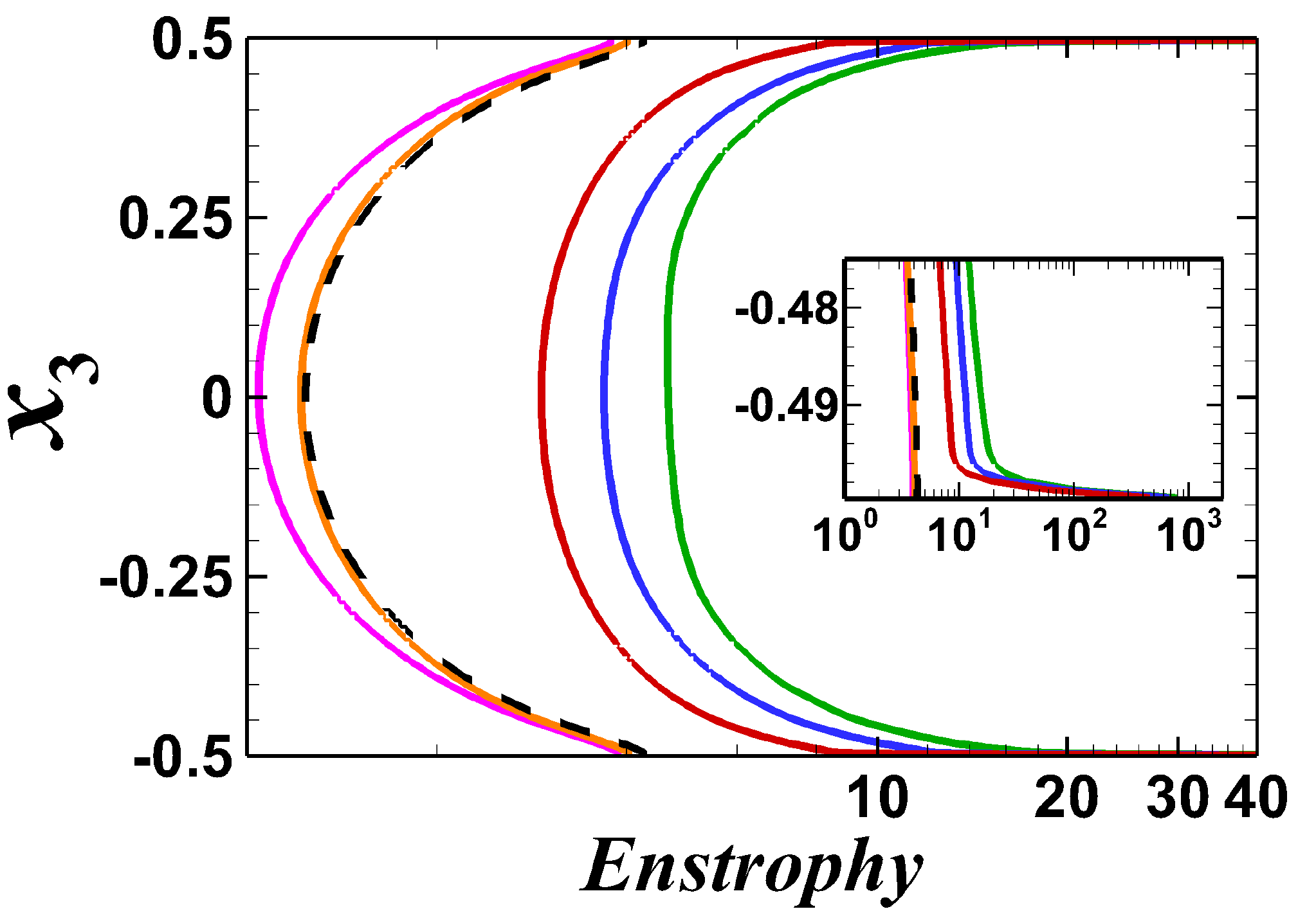}
(b) \includegraphics[width=0.448\linewidth,trim={1.8cm 6.2cm 0cm 9cm},clip]{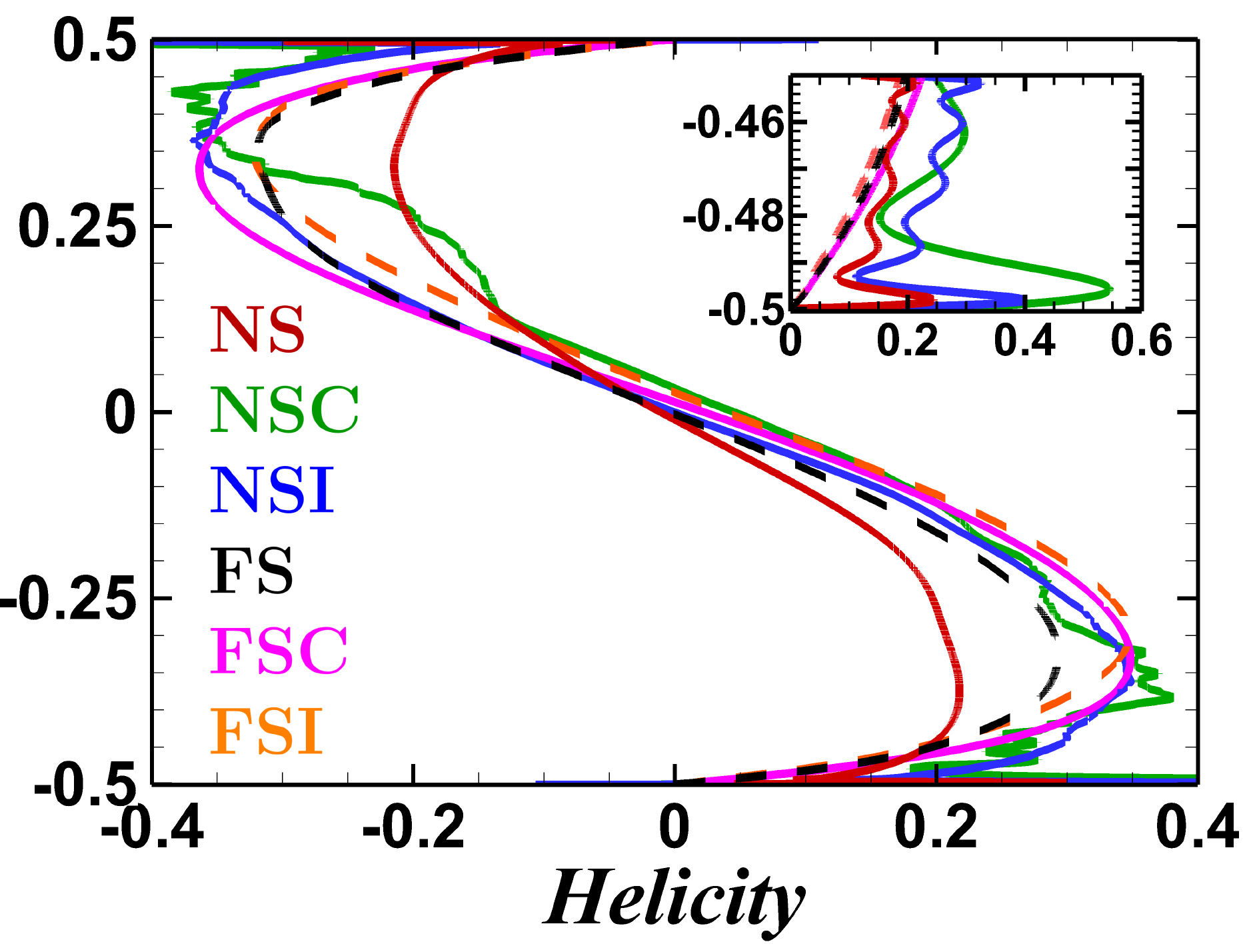}\\
(c) \includegraphics[width=0.475\linewidth,trim={0.2cm 6.2cm 0cm 9cm},clip]{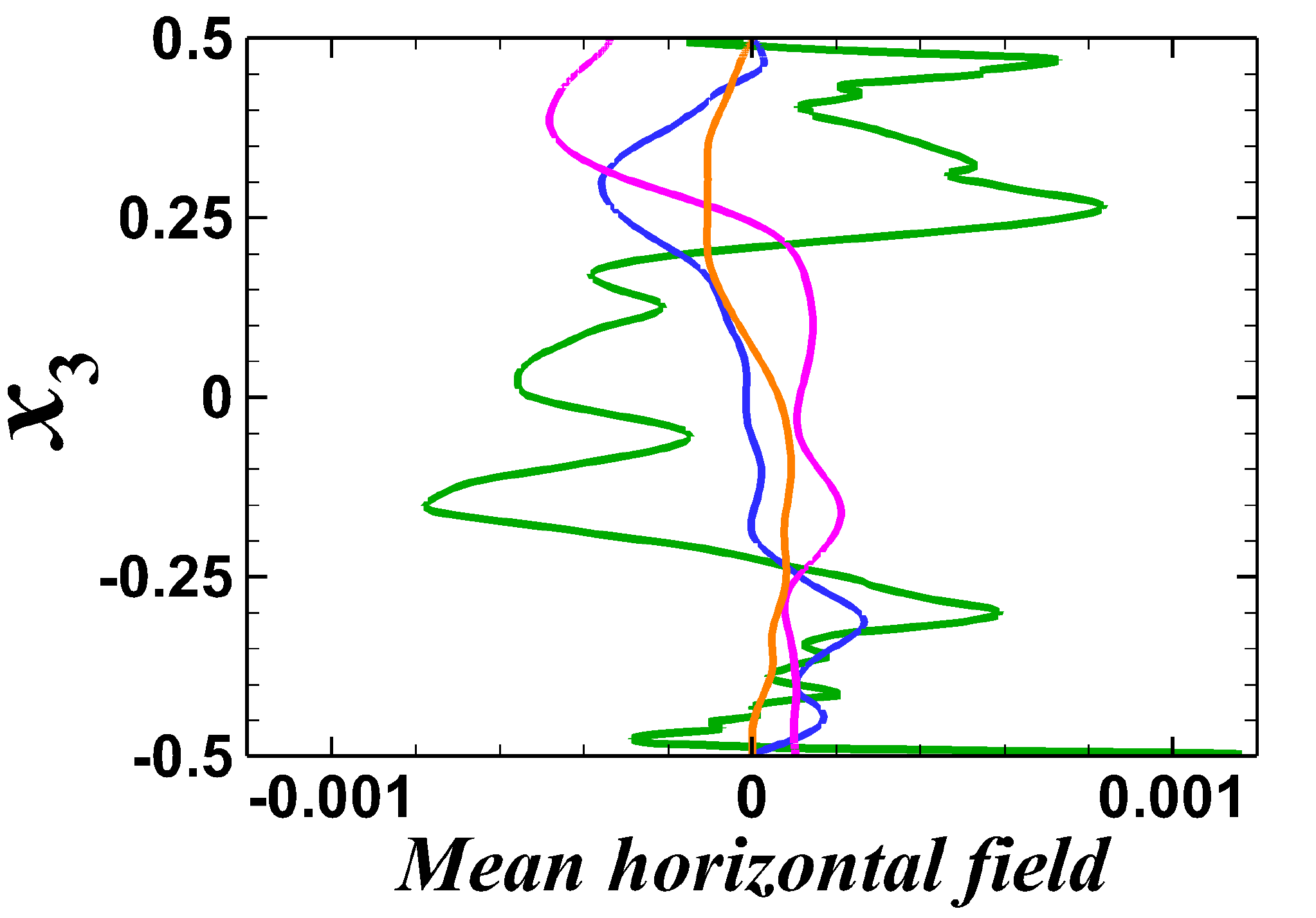}
(d) \includegraphics[width=0.455\linewidth,trim={1.8cm 6.2cm 0cm 5cm},clip]{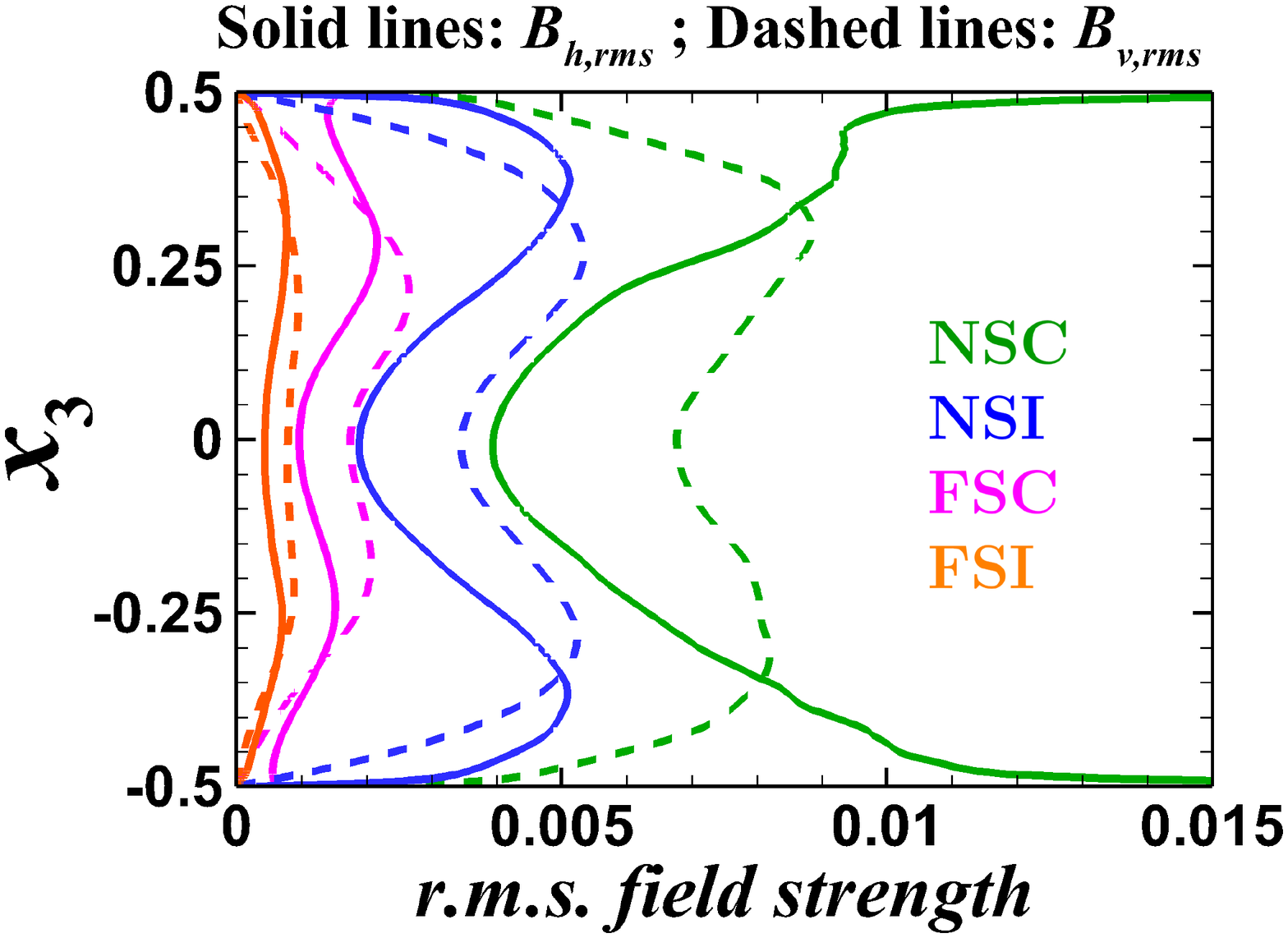}

\caption{Vertical variation of \textit{r.m.s.} quantities: (\textit{a}) Enstrophy, (\textit{b}) Helicity, (\textit{c}) mean magnetic field, (\textit{d}) \textit{r.m.s.} magnetic field at $\mathcal{R}=3$. All quantities are averaged in time and in the horizontal directions.}
\label{fig:mag}
\end{figure*}

Apart from the velocity and thermal fields, we look into the effect of boundary conditions on the enstrophy, relative helicity and the magnetic field. The vertical variations of horizontally averaged enstrophy, relative helicity, mean and \textit{r.m.s.} magnetic field strengths are depicted in figure \ref{fig:mag}a,b,c and d respectively. The enstrophy, a measure of the strength of the vortical elements in the flow, indicates the extent to which they can deform the magnetic field lines and therefore play a key role in deciding the local Lorentz force magnitude. Vorticity fluctuations are enhanced in the bulk, as depicted by the enstrophy ($E_{\omega}=1/2\ \overline{\omega_i\omega_i}$, with more than two orders of magnitude jump near the boundaries (see the inset in figure \ref{fig:mag}a). The reason for this increase in enstrophy is the increased strength of the vortices due to Ekman pumping near the wall. The presence of energetic vortices near the wall may significantly alter the boundary layer dynamics and the associated heat transfer characteristics of a dynamo compared to the same without the presence of Ekman layer with free-slip boundaries \citep{naskar_2021}. Another important quantity is the kinetic helicity of the flow, which can induce large-scale mean fields in a dynamo \citep{tilgner_2012}. The relative kinetic helicity, $\mathcal{H}_r=\overline{u_{i}\omega_{i}}/2\left(KE_{\omega}\right)^{1/2}$, exhibit the well-known sinusoidal distribution in the vertical direction, as expected in rotating convection, with negative and positive helicity dominating in the bottom and top halves of the domain respectively \citep{cattaneo_2006,schmitz_2010}. Helicity is enhanced by the presence of the wall, where the thermal plumes departing from the boundary layer towards the bulk are spun up by Coriolis force, due to Ekman pumping \citep{schmitz_2010}. This phenomenon results in a strong correlation between local velocity and vorticity that leads to a peak of relative kinetic helicity near the wall, as shown in the inset in figure \ref{fig:mag}b. Additionally, we look into the effect of boundary conditions on the strength and structure of the magnetic field produced by the dynamos. Horizontally averaged mean magnetic field is plotted in figure \ref{fig:mag}c, which illustrates the dependence on magnetic boundary conditions, even in the bulk. NSC conditions lead to the highest mean-field magnitude among all the cases. It should be noted here that the vertical component of the mean magnetic field, $\Bar{B}_3$, is identically zero by the definition of averages and the solenoidal field condition so that the mean-field remains horizontal. However, the fluctuating part of the magnetic field is three-dimensional, with all components non-zero except at the wall. The \textit{r.m.s.} value of the fluctuating horizontal and vertical magnetic fields are plotted in figure \ref{fig:mag}d. For $\mathcal{R}=3$, the fluctuating magnetic field is approximately one order of magnitude stronger than the mean magnetic field. However, lower thermal forcing can generate strong, large-scale mean magnetic fields as reported in earlier studies \citet{stellmach_2004,tilgner_2012,naskar_2021}. For insulated boundary conditions, the horizontal and vertical components of the magnetic field are zero at the boundaries, whereas for conducting conditions, the horizontal components remain non-zero at the boundaries. Dynamos with no-slip boundary conditions lead to higher \textit{r.m.s.} field strength than that of free-slip boundaries. For NSC conditions, a large horizontal \textit{r.m.s.} field magnitude can be observed near the boundaries. The stretching of the magnetic field lines by the strong vortices near the wall results in an increase in the \textit{r.m.s.} field strength. As the magnetic field has to remain parallel to a perfectly conducting surface, it remains trapped near the walls, leading to a build-up of magnetic field strength \citep{stpierre_1993}. Overall structure and magnitude of the magnetic field, both in bulk and near the boundaries, are strongly dependent on the combination of kinematic and magnetic boundary conditions for all $\mathcal{R}$. The traditional Elsasser number, $\Lambda_{tr}=\sigma B_\tau^2/\rho\Omega$, where $B_\tau=(B_{1,rms}^2+B_{2,rms}^2+B_{3,rms}^2)^{1/2}$ provides a non-dimensional measure of the \textit{r.m.s} magnetic field strength, as reported in tables \ref{tab:nsc}-\ref{tab:fsi}. The \textit{r.m.s} magnetic field monotonically increases with increasing thermal forcing for all boundary conditions. However, the mean-field strength is found to decrease with increasing $\mathcal{R}$ \citep{naskar_2021}, indicating a shift towards small-scale dynamo action with increasing thermal forcing \citep{tilgner_2012,tilgner_2014}. 

\subsection{Force balance}\label{sec:forces}

\begin{figure*}
\centering
(a) \includegraphics[width=0.48125\linewidth,trim={0.2cm 8.1cm 0cm 7cm},clip]{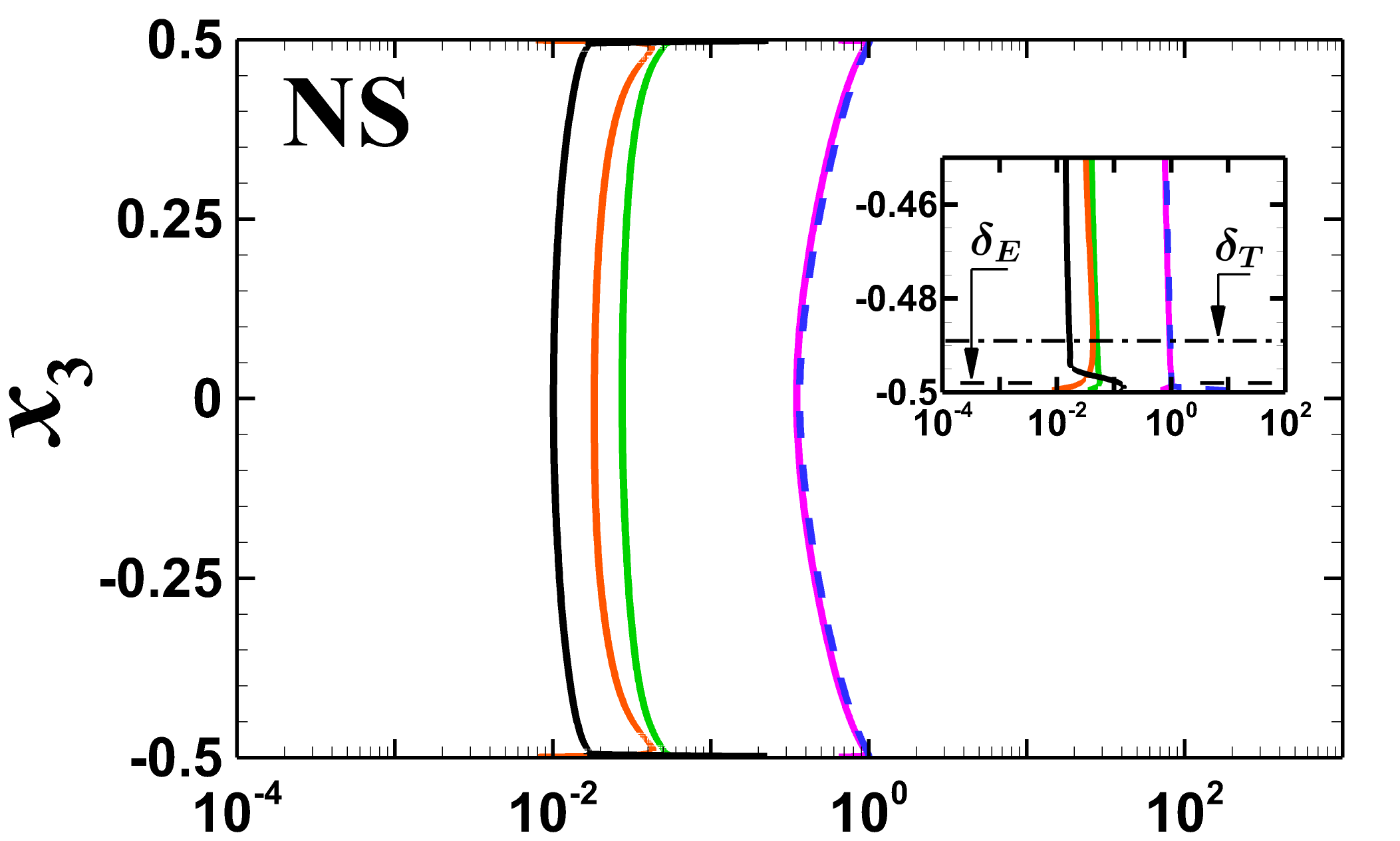}
(b) \includegraphics[width=0.44875\linewidth,trim={1.8cm 8.1cm 0cm 7cm},clip]{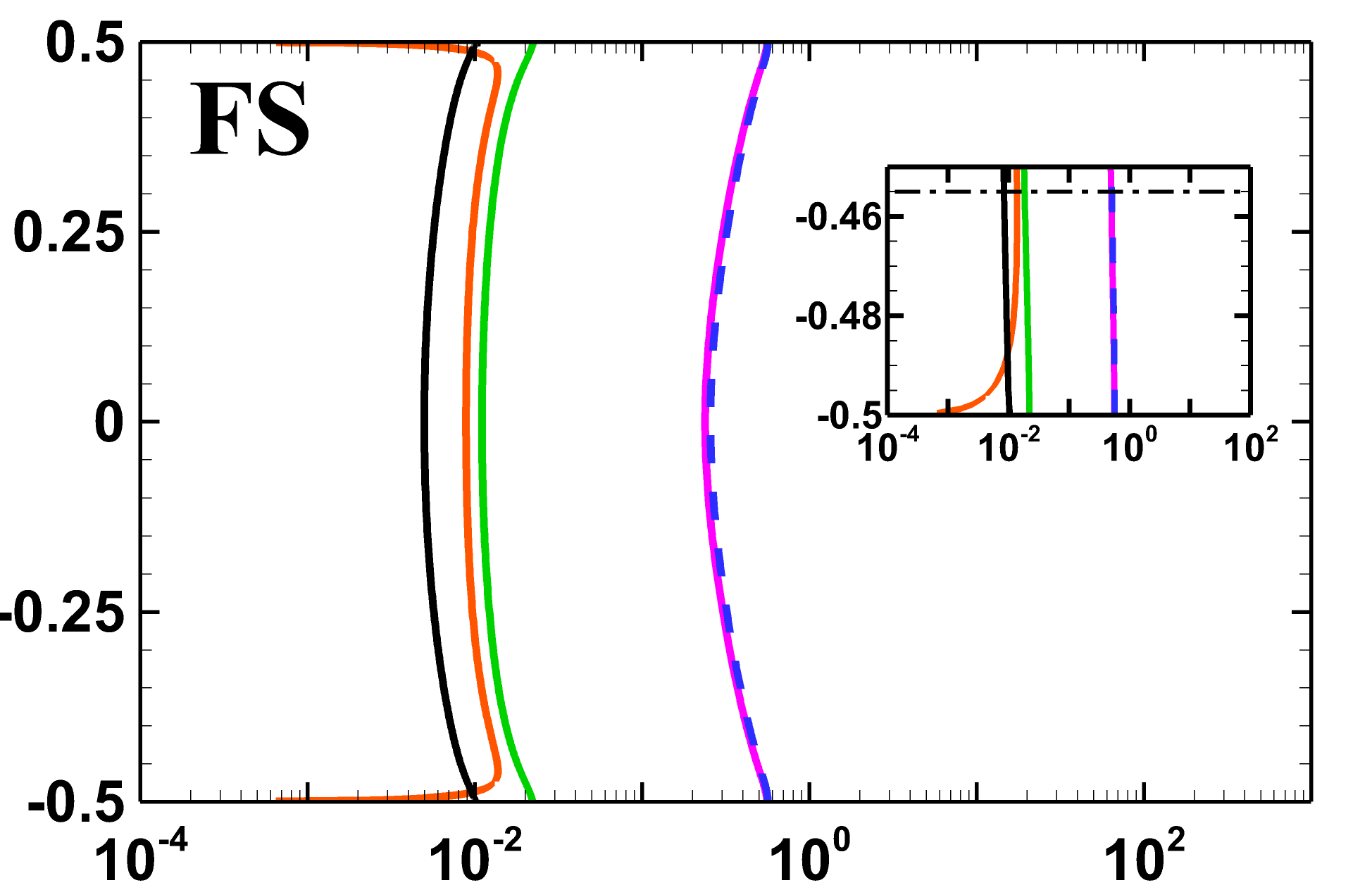}\\
(c) \includegraphics[width=0.48125\linewidth,trim={0.2cm 8.1cm 0cm 7cm},clip]{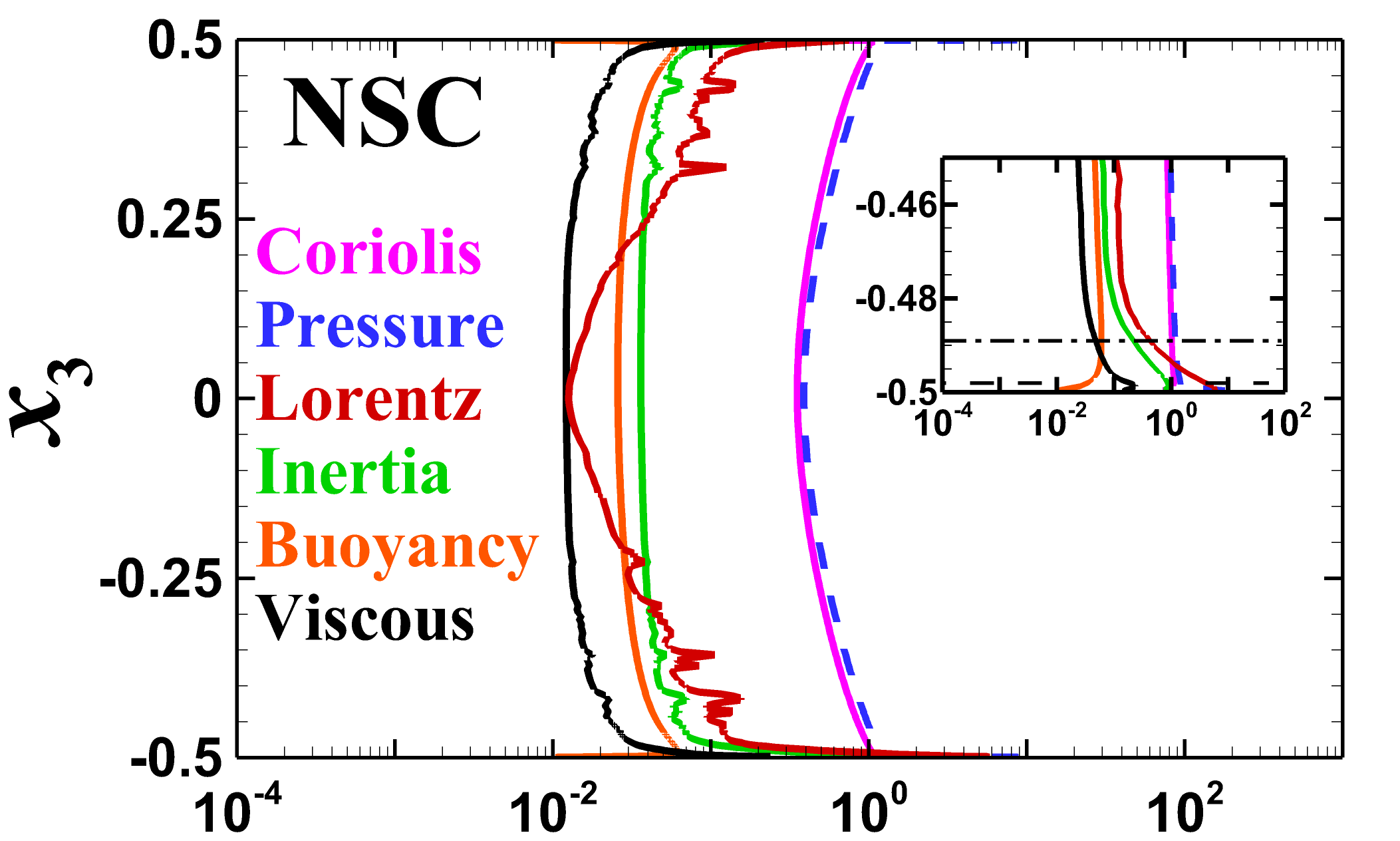}
(d) \includegraphics[width=0.44875\linewidth,trim={1.8cm 8.1cm 0cm 7cm},clip]{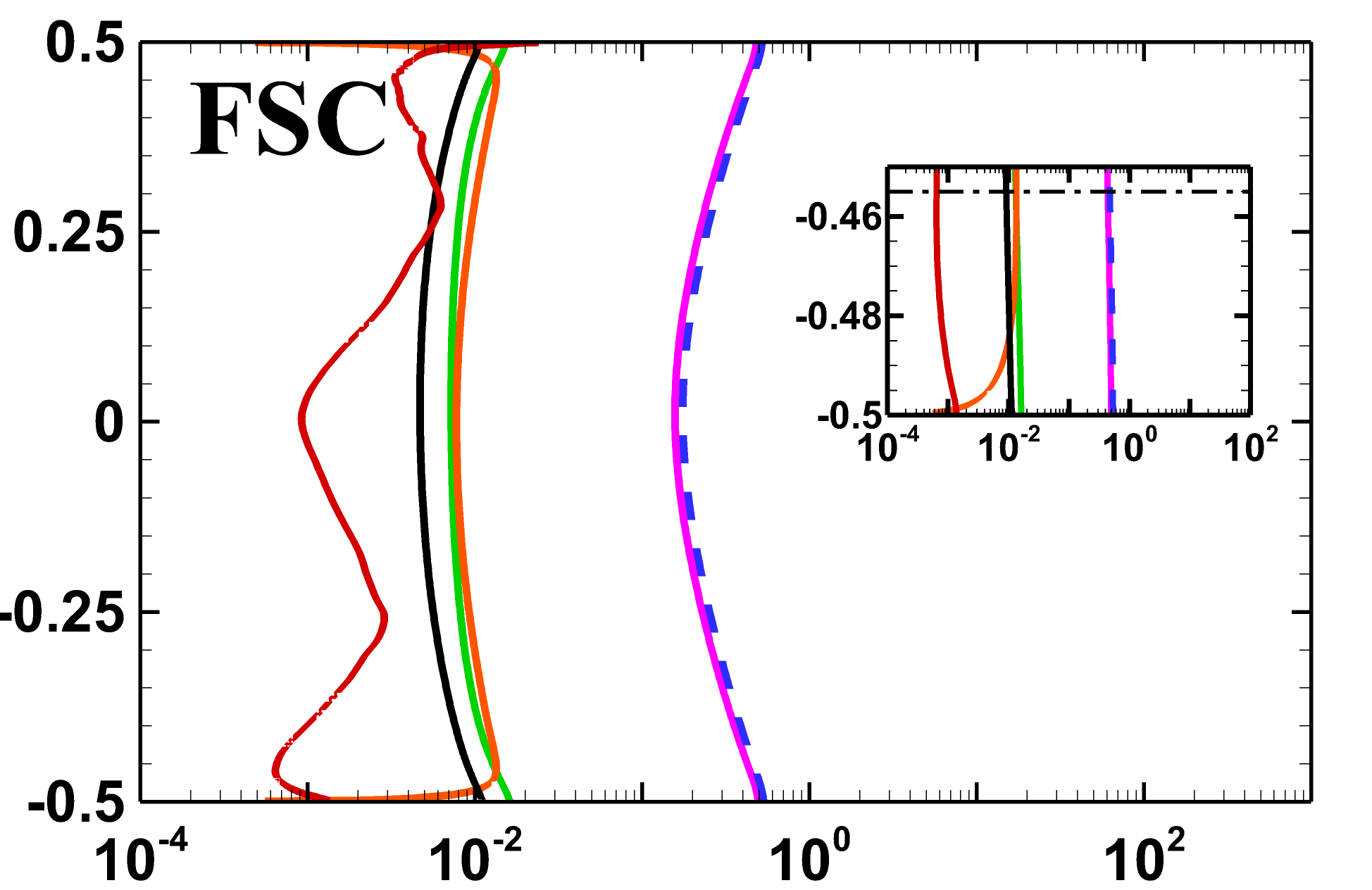}\\
(e) \includegraphics[width=0.48125\linewidth,trim={0.2cm 7cm 0cm 7cm},clip]{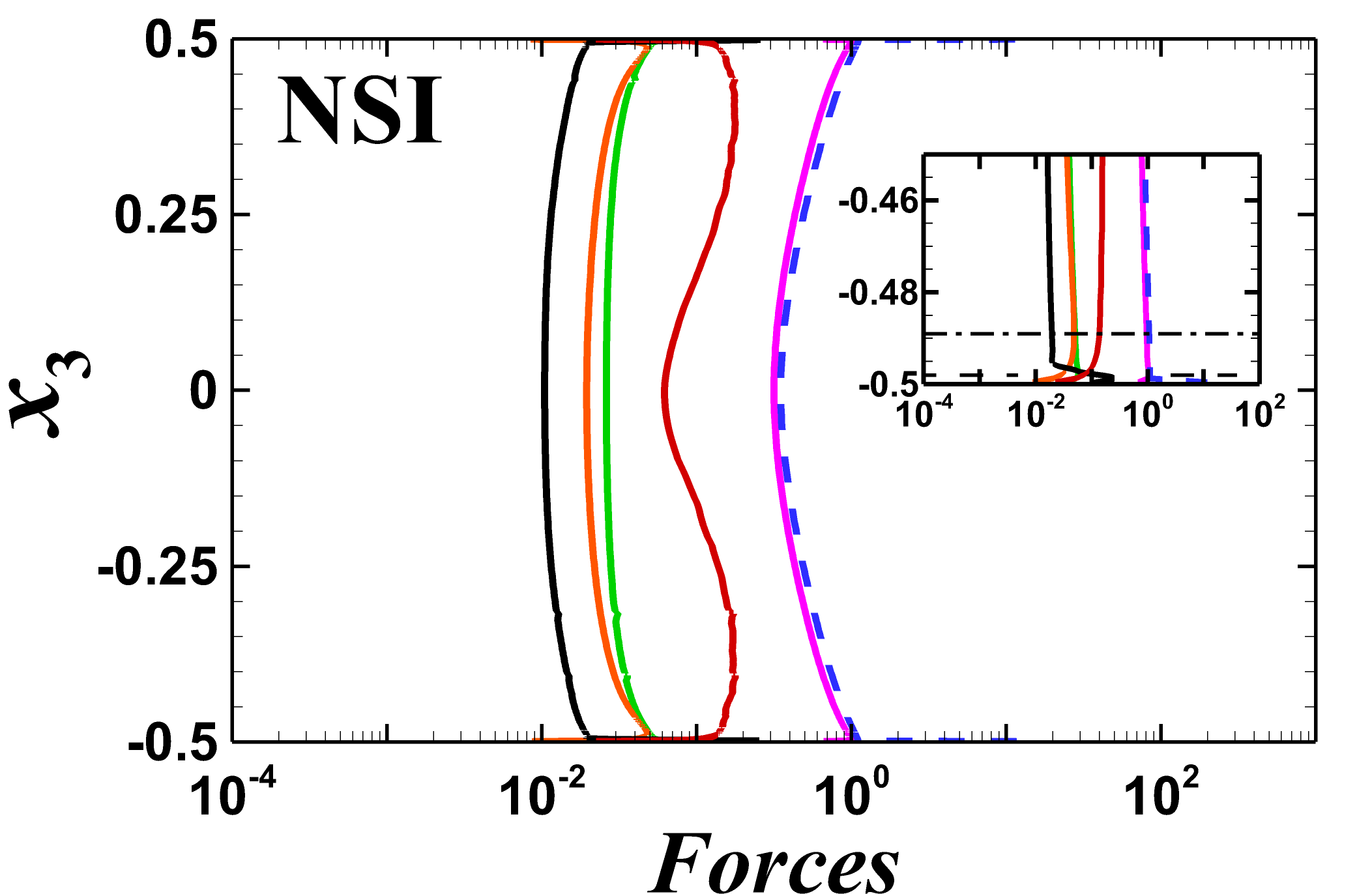}
(f) \includegraphics[width=0.44875\linewidth,trim={1.8cm 7cm 0cm 7cm},clip]{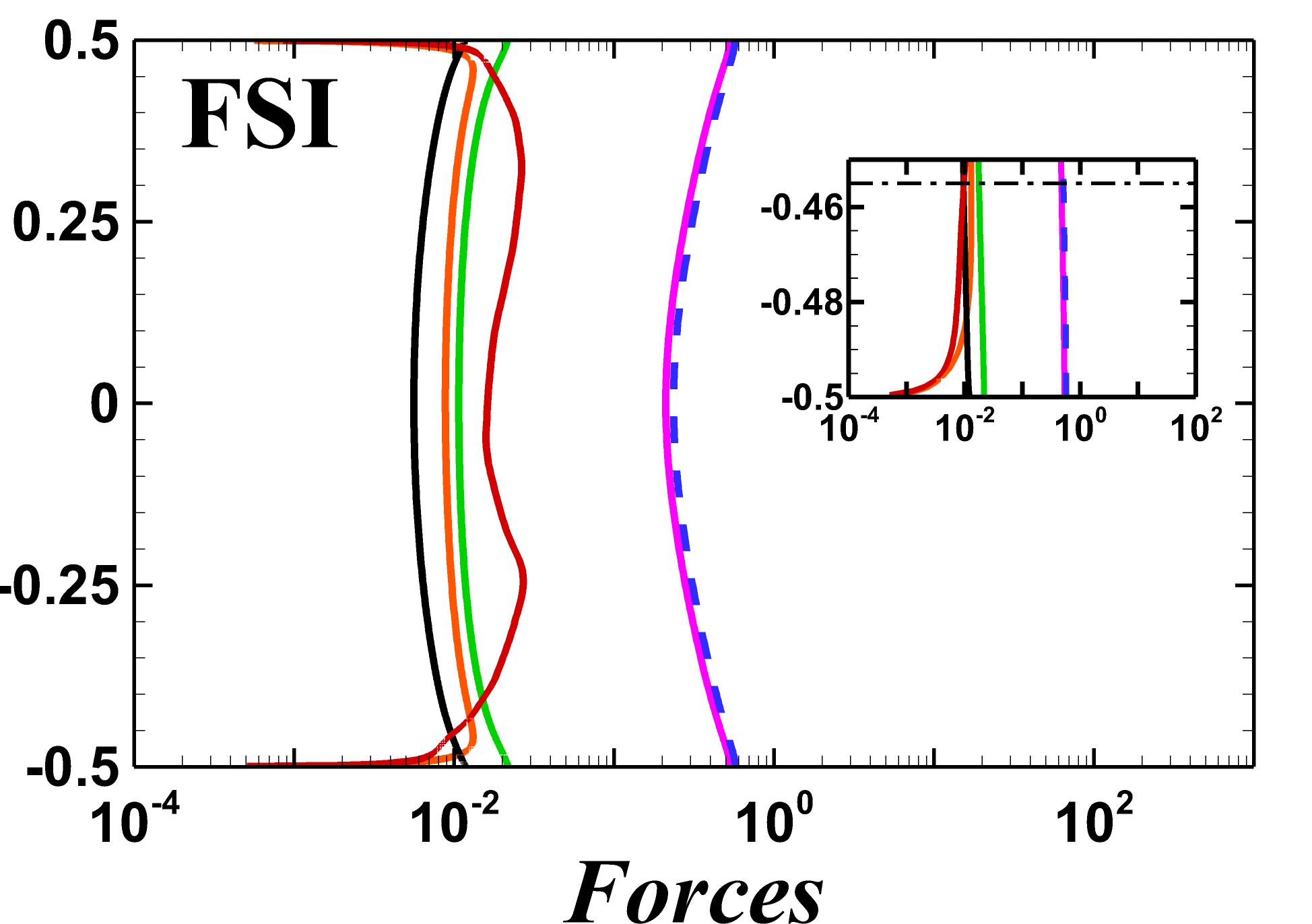}
\caption{Vertical variation of forces  for (\textit{a}) NS, (\textit{b}) FS, (\textit{c}) NSC ,(\textit{d}) FSC ,(\textit{e}) NSI, (\textit{f}) FSI cases at $\mathcal{R}=3$. The horizontally averaged force distribution is shown in the bulk and near the bottom plate(inset).}
\label{fig:forces}
\end{figure*}

In this section, we report a comparative study of dynamical balances with different kinematic and magnetic boundary conditions at $\mathcal{R}=3$. Figure \ref{fig:forces} shows the vertical variation of horizontally averaged forces, evaluated from the \textit{r.m.s.} values of each terms in the momentum equation \ref{eqn:momentum_nd} \citep{yan_2021,guzman_2021} . The first and second columns in this figure present the forces for no-slip and free-slip boundary conditions, respectively at $\mathcal{R}=3$. Here, the non-magnetic convection results (first row) are used as a reference to interpret results for dynamo simulations with perfectly conducting (second row) and insulated (third row) boundary conditions. At $\mathcal{R}=3$, we get the thermal and Ekman boundary layer thickness as $\delta_T=0.011$ and $\delta_E=0.002$ for the no-slip cases, whereas the thermal layer thickness increases to $\delta_T=0.045$ for free-slip condition. The near-wall regions are magnified in the insets, with the velocity and thermal boundary layer edges marked by black horizontal dashed and dash-dot lines respectively in figure \ref{fig:forces}a. For all the cases shown in this figure, the leading order balance between Coriolis and pressure force indicates a geostrophic state in the bulk. For non-magnetic rotating convection, geostrophic balance in the bulk has been confirmed by DNS \citep{guzman_2021}, apart from reduced-order models in the rapidly rotating limit, $E\xrightarrow{}0$ \citep{julien_2012b}. Departure from the geostrophic state due to the other forces, which constitute a lower order quasi-geostrophic balance, makes turbulent convection possible. In the non-magnetic simulations (NS and FS) in figures \ref{fig:forces}a and b, the geostrophic and quasi-geostrophic forces behave similarly except near the boundaries (see insets), where viscous force break the geostrophic balance and dominate the other quasi-geostrophic forces (inertia and buoyancy) in the Ekman layer near the plates with the no-slip boundary condition.\\

For dynamo simulations, the Lorentz force exerted by the magnetic field on the fluid enters the quasi-geostrophic balance. For the NSC case as shown in figure \ref{fig:forces}c, the Lorentz force is minimum at the mid-plane, which increases towards the walls to dominate the quasi-geostrophic balance for $x_3\leq-0.25$. Inside the thermal boundary layer, the Lorentz force increases to the same order of magnitude as the Coriolis force, and eventually becomes the highest force at the wall. This is reflected by the value of local Elsasser number, $\Lambda_T$ (the ratio of the \textit{r.m.s.} magnitudes of the Lorentz and the Coriolis forces at the edge of the thermal boundary layer, calculated from the horizontally averaged variation of the two forces), as presented in the tables \ref{tab:nsc}-\ref{tab:fsi}. This increase in the Lorentz force at the thermal boundary layer edge leads to local magnetorelaxation of the thermal boundary layer, which results in increased turbulence and heat transport \citep{naskar_2021}. Similar behavior of the Lorentz force near the boundary was found in the range $\mathcal{R}=3-5$ for the NSC cases. However, no such enhancement of Lorentz force is found in the dynamo simulations with any other combinations of boundary conditions. For the NSI case in figure \ref{fig:forces}e, the Lorentz force is at least five times higher than other quasi-geostrophic forces both in the bulk and inside the thermal layer. It decreases near the Ekman layer, where viscous force dominates. The volume-averaged ratio of the Lorentz and the Coriolis forces is also presented as $\Lambda_V$ in tables \ref{tab:nsc}-\ref{tab:fsi}. This volume-averaged Elsasser number reaches a maximum near $\mathcal{R}=4$ for the NSI case in table \ref{tab:nsi}. The Lorentz force inside the thermal layer is one order of magnitude smaller than other quasi-geostrophic forces for FSC case in figure \ref{fig:forces}d, making the near-wall balance similar to non-magnetic rotating convection (FS). Though the Lorentz force is the highest ageosptrophic force in the bulk for the FSI cases in figure \ref{fig:forces}f, it remains the smallest force inside the thermal boundary layer, again making the balance similar to the FS case near the walls. These results illustrate the dependence of the dynamical balance of the dynamo on the imposed boundary conditions, especially near the boundary. Also, the dynamo simulations with no-slip conditions exhibit distinctive balance compared to the  non-magnetic convection inside the thermal boundary layer. In contrast, the dynamical balance in the thermal layer is similar to rotating convection for free-slip boundary conditions with negligible contribution from the Lorentz force.\\    

\begin{figure*}
\centering
(a) \includegraphics[width=0.48125\linewidth,trim={0.2cm 8.1cm 0cm 7cm},clip]{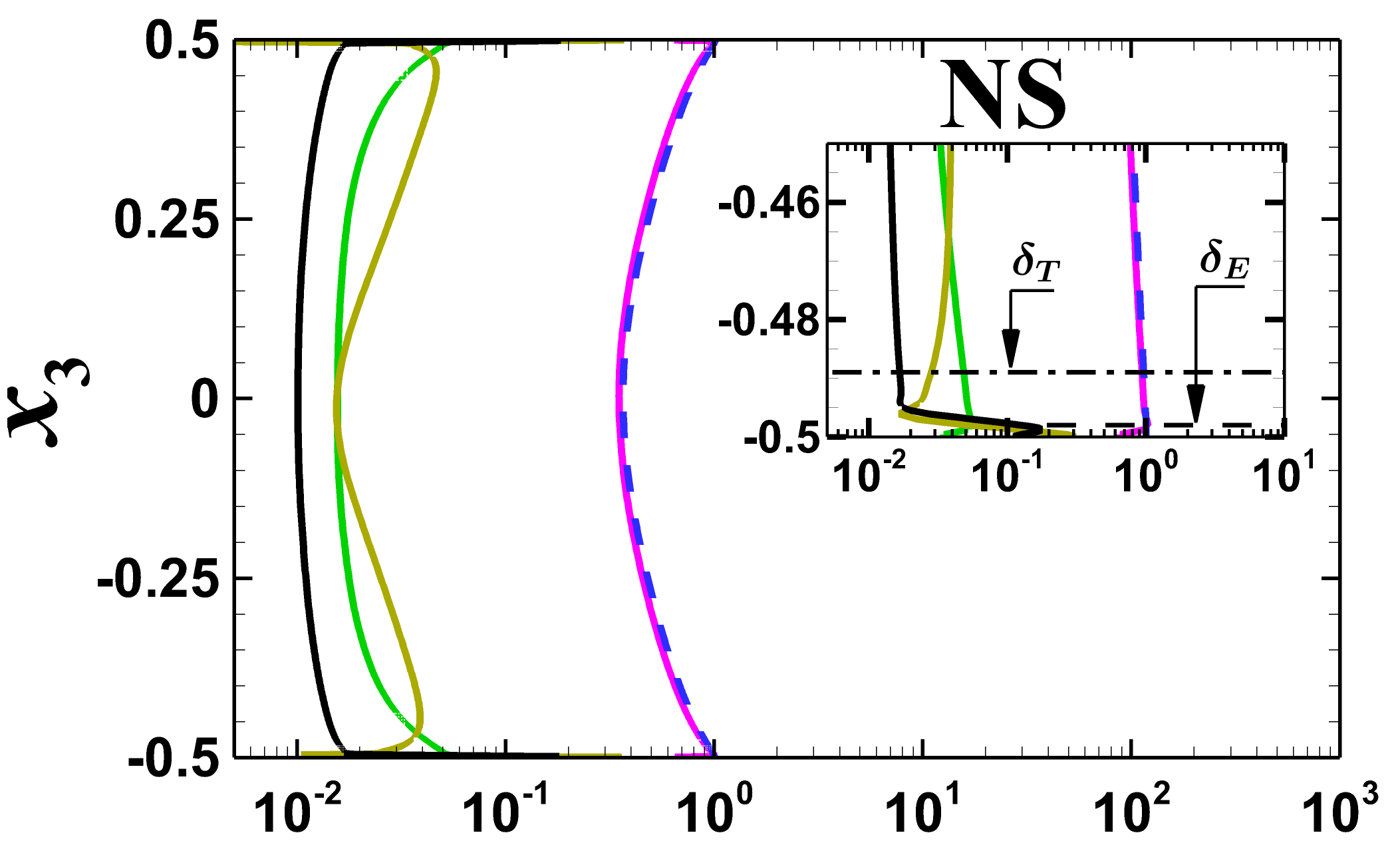}
(b) \includegraphics[width=0.44875\linewidth,trim={1.8cm 8.1cm 0cm 7cm},clip]{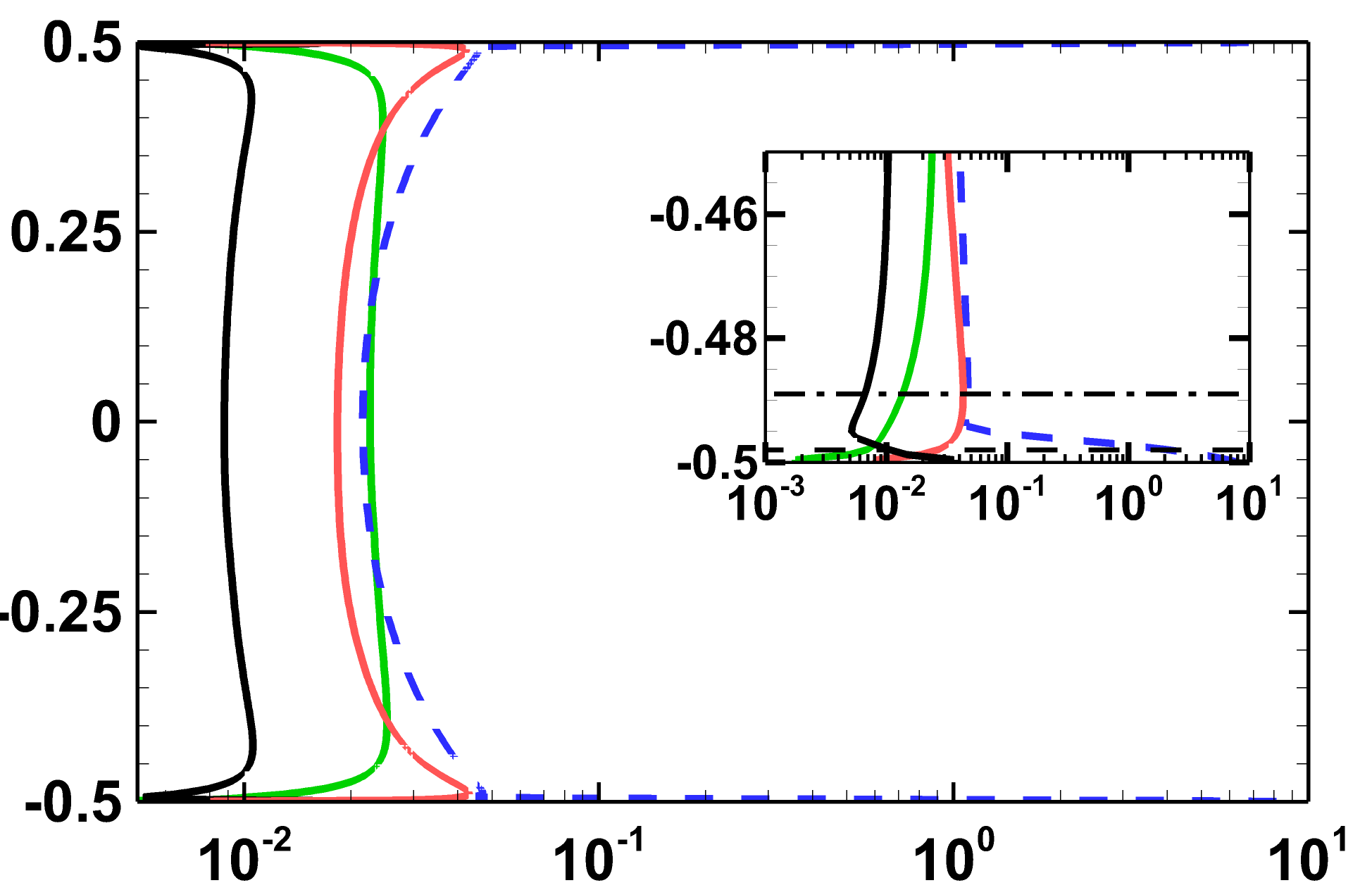}\\
(c) \includegraphics[width=0.48125\linewidth,trim={0.2cm 6.5cm 0cm 7cm},clip]{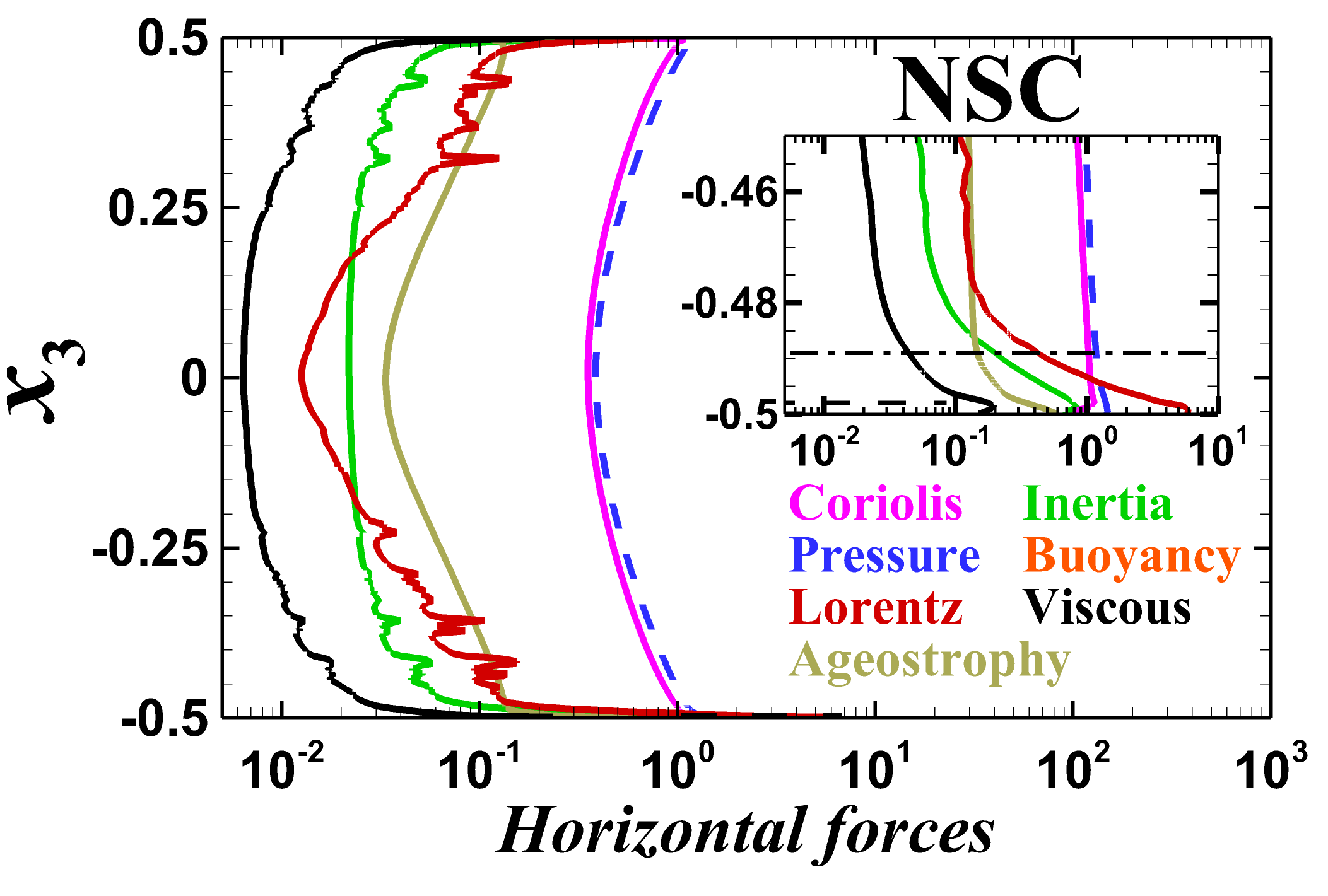}
(d) \includegraphics[width=0.44875\linewidth,trim={1.8cm 6.5cm 0cm 7cm},clip]{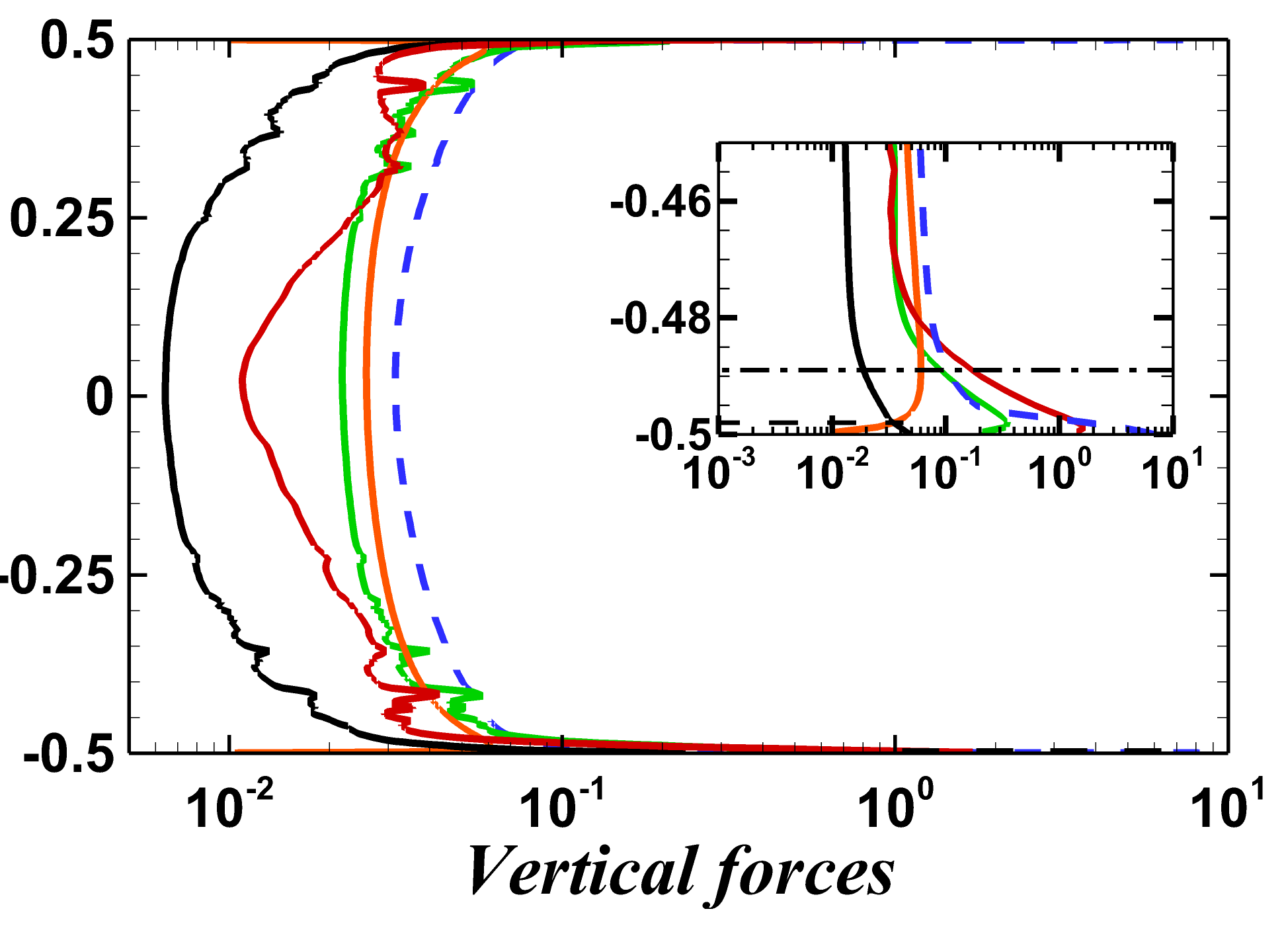}
\caption{Vertical variation of force components for (\textit{a},\textit{b}) NS and (\textit{c},\textit{d}) NSC cases. Horizontal component of the forces $F_h=\sqrt{F^2_{x_1}+F^2_{x_2}}$ are shown in left column while their vertical components ($F_{x_3}$) are in the right column. Horizontally averaged force distribution is shown here at $\mathcal{R}=3$ both in the bulk, and near the bottom plate (inset)}
\label{fig:force_comp}
\end{figure*}

The force balance in the NSC case at $\mathcal{R}=3$ exhibits interesting features near the boundary layer as shown in figure \ref{fig:forces}. This case can be analyzed in more details from figure \ref{fig:force_comp}, where the horizontal and vertical balances are illustrated and compared with the NS case. As the rotation axis is aligned with the vertical, the Coriolis force acts only in the horizontal planes, whereas the buoyancy force is purely vertical. In the non-magnetic case, the leading order horizontal balance is geostrophic between the Coriolis and pressure forces as depicted in figure \ref{fig:force_comp}a. The ageostrophy, defined by the difference between the two forces, is balanced by inertia and viscous forces. Near the walls, there is a sharp increase in the ageostrophy, balanced by viscous force, indicating loss of geostrophic balance inside the viscous boundary layers, as seen in the inset of figure \ref{fig:force_comp}a \citep{guzman_2021}. For the vertical forces in figure \ref{fig:force_comp}b, the vertical pressure gradient is balanced by the inertia and buoyancy forces in the bulk where viscous forces remain small.\\

\begin{figure*}
\centering
(a) \includegraphics[width=0.48125\linewidth,trim={0.2cm 7cm 0cm 7cm},clip]{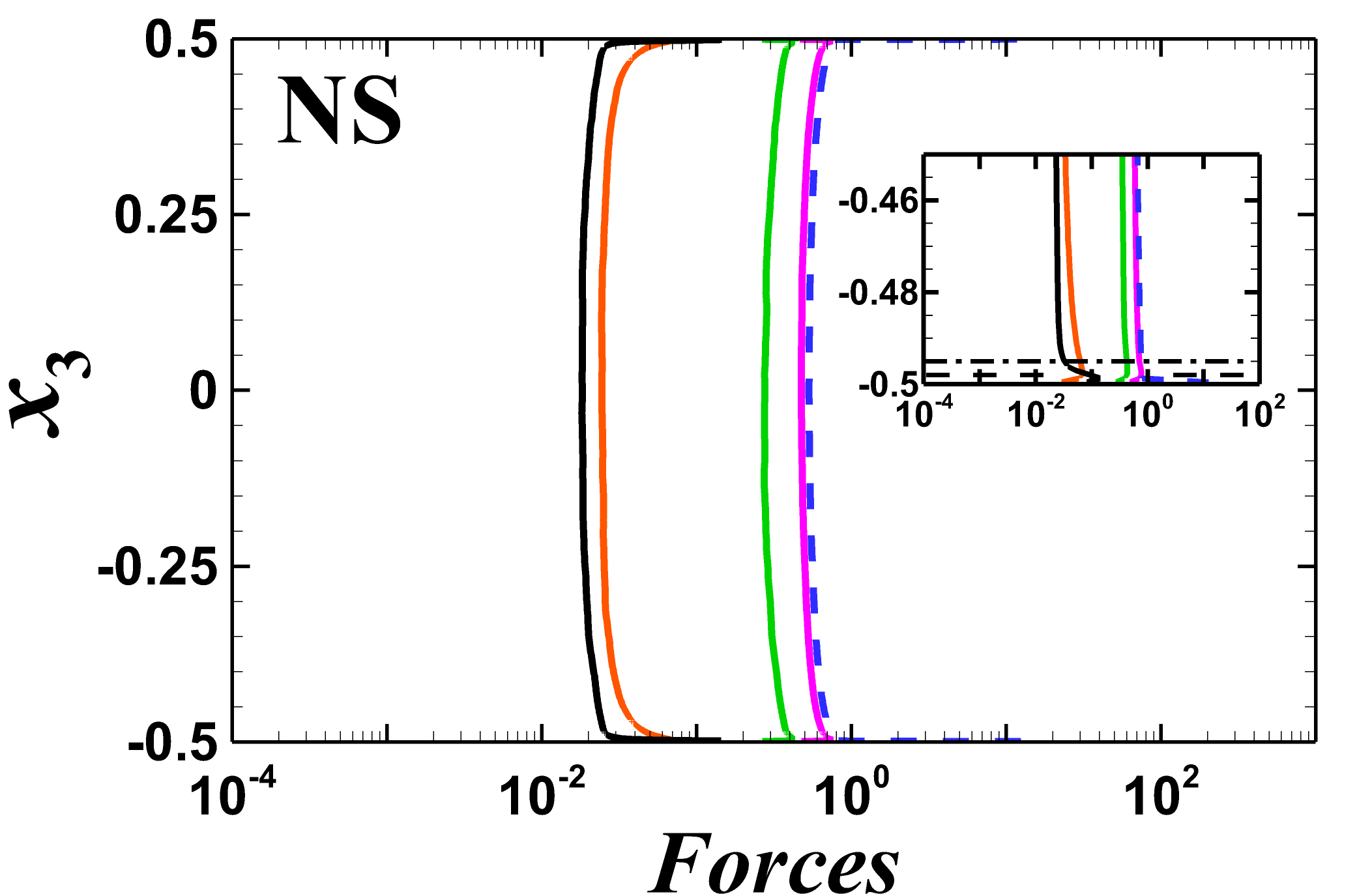}
(b) \includegraphics[width=0.44875\linewidth,trim={1.8cm 7cm 0cm 7cm},clip]{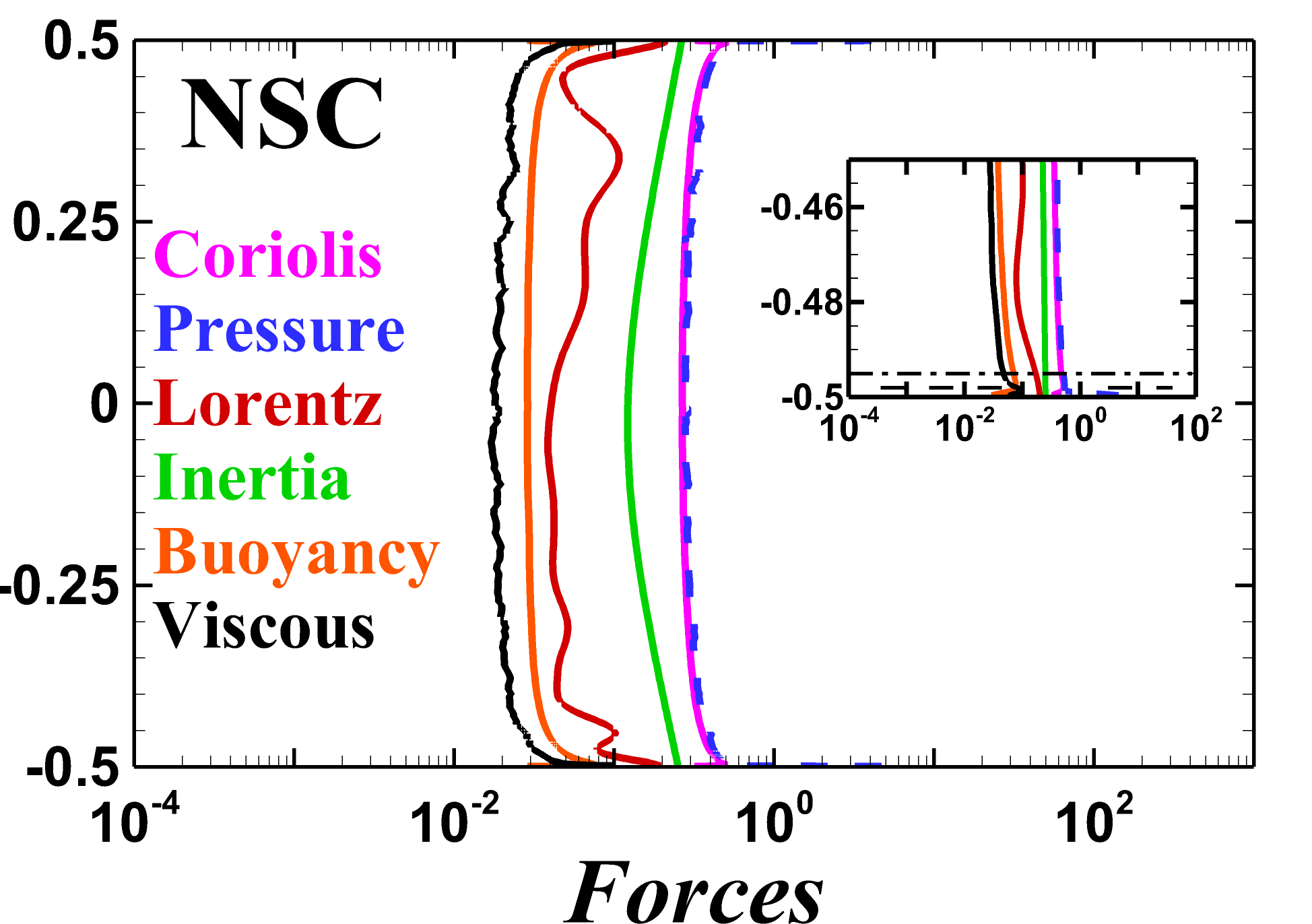}\\
\caption{Vertical variation of forces  for (\textit{a}) NS, (\textit{b}) NSC cases at $\mathcal{R}=20$. The horizontally averaged force distribution is shown in the bulk and near the bottom plate(inset).}
\label{fig:forces20}
\end{figure*}

In the NSC case, the leading order balance remains to be geostrophic in the horizontal direction as presented in figure \ref{fig:force_comp}c. However, the ageostrophy is much higher than the NS case due to magnetorelaxation of the Taylor-Proudman theorem in the presence of Lorentz force. The ageostrophic component is balanced by the Lorentz and the inertia forces in the bulk. In the vertical direction, the pressure force is balanced by Lorentz, buoyancy, and inertia forces as shown in figure \ref{fig:force_comp}d. Notice that the horizontal components of the forces are always higher than the vertical with increased force magnitudes near the boundaries for all the cases. In the Ekman layer, the flow velocities gradually come to zero resulting in an increase in pressure at the boundary. It should be noted here that the details of the force balance, especially the role of inertia in the force balance, depend on $\mathcal{R}$. With increase in $\mathcal{R}$, the inertia force eventually dominates the quasi-geostrophic balance as observed for $\mathcal{R}=20$ in figure \ref{fig:forces20}a and b for NS and NSC cases respectively. This increase in inertia force with thermal forcing corroborates the finding of \citet{guzman_2021}, and a common feature for all boundary conditions. The local magnetorelaxation of the rotational constraint in the thermal boundary layer, as a mechanism for increase in heat transfer in NSC case compared to NS case, becomes gradually ineffective with increasing $\mathcal{R}$. This is because of the increased inertia force, rather than the Lorentz force breaks the Taylor-Proudman constraint for both non-magnetic and dynamo simulations, irrespective of boundary conditions.       

\subsection{Energy Budget}\label{sec:budget}
%, with variation near the bottom plate magnified in the insets to elucidate the near wall dynamics.

\begin{figure*}
\centering
(a) \includegraphics[width=0.463\linewidth,trim={0cm 1.2cm 0cm 0cm},clip]{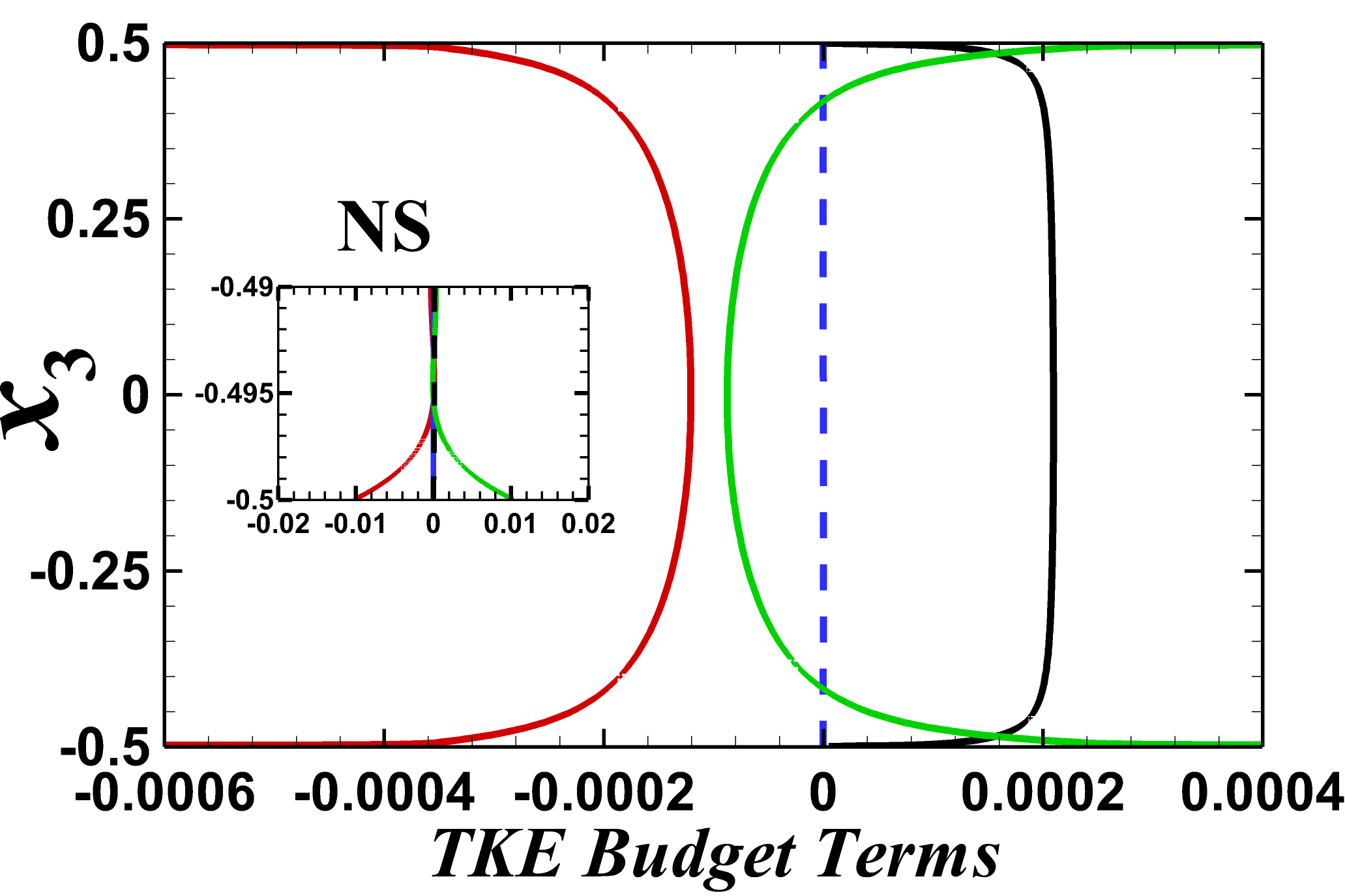}
(b) \includegraphics[width=0.467\linewidth,trim={0cm 1.2cm 0cm 0cm},clip]{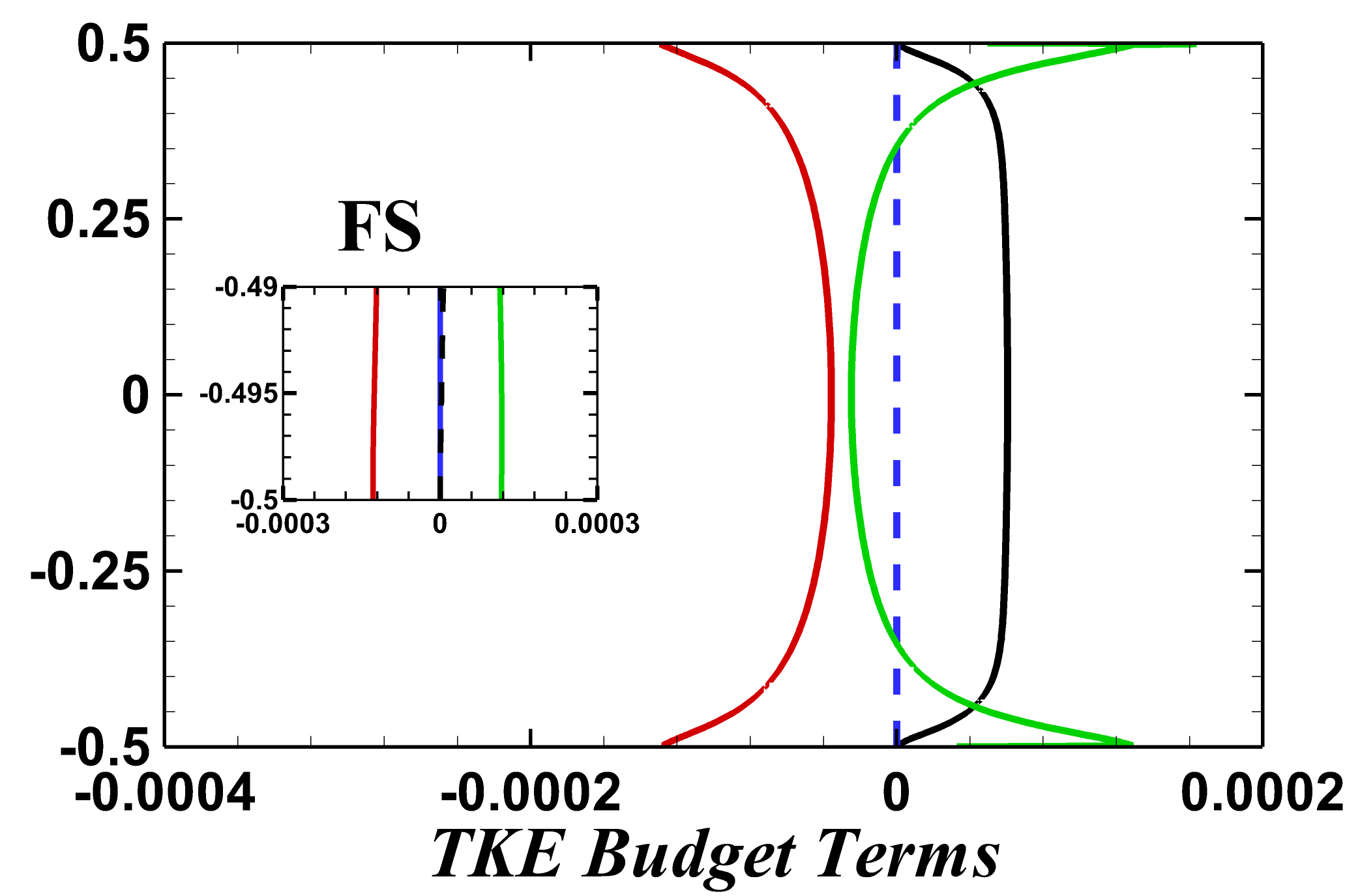}\\
(c) \includegraphics[width=0.463\linewidth,trim={0cm 1.2cm 0cm 0cm},clip]{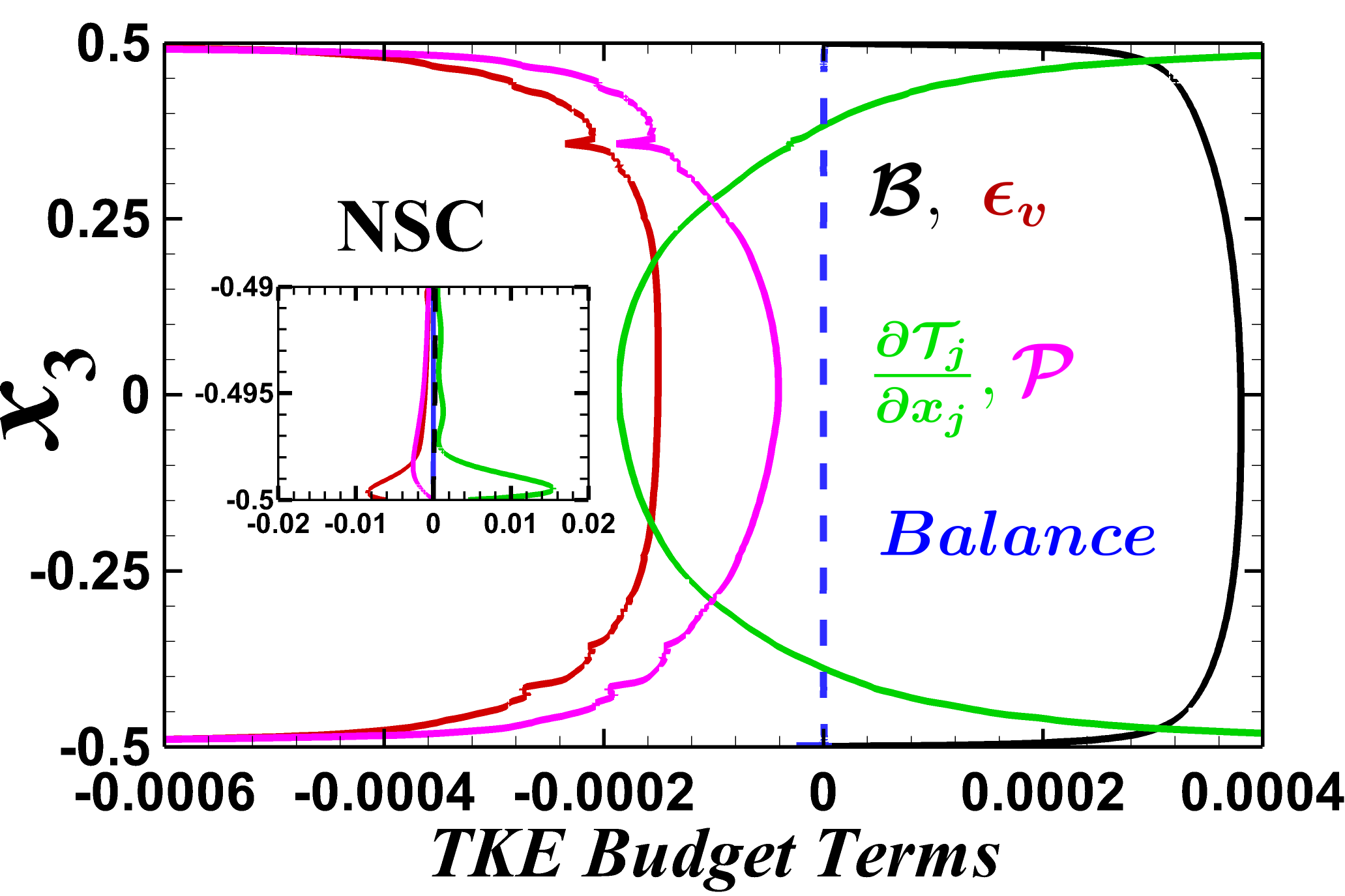}
(d) \includegraphics[width=0.467\linewidth,trim={0cm 1.2cm 0cm 0cm},clip]{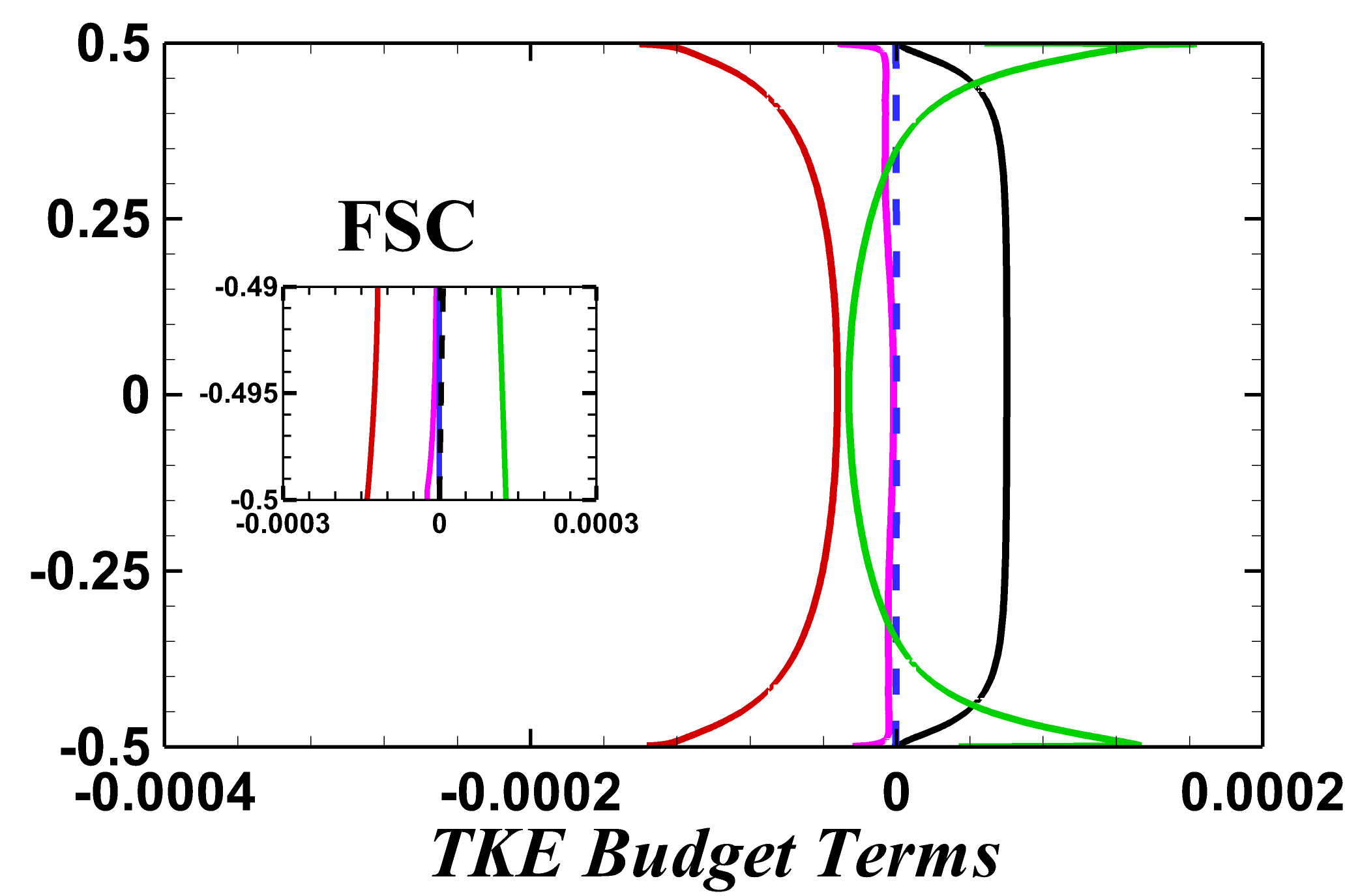}\\
(e) \includegraphics[width=0.463\linewidth,trim={0cm 0cm 0cm 0cm},clip]{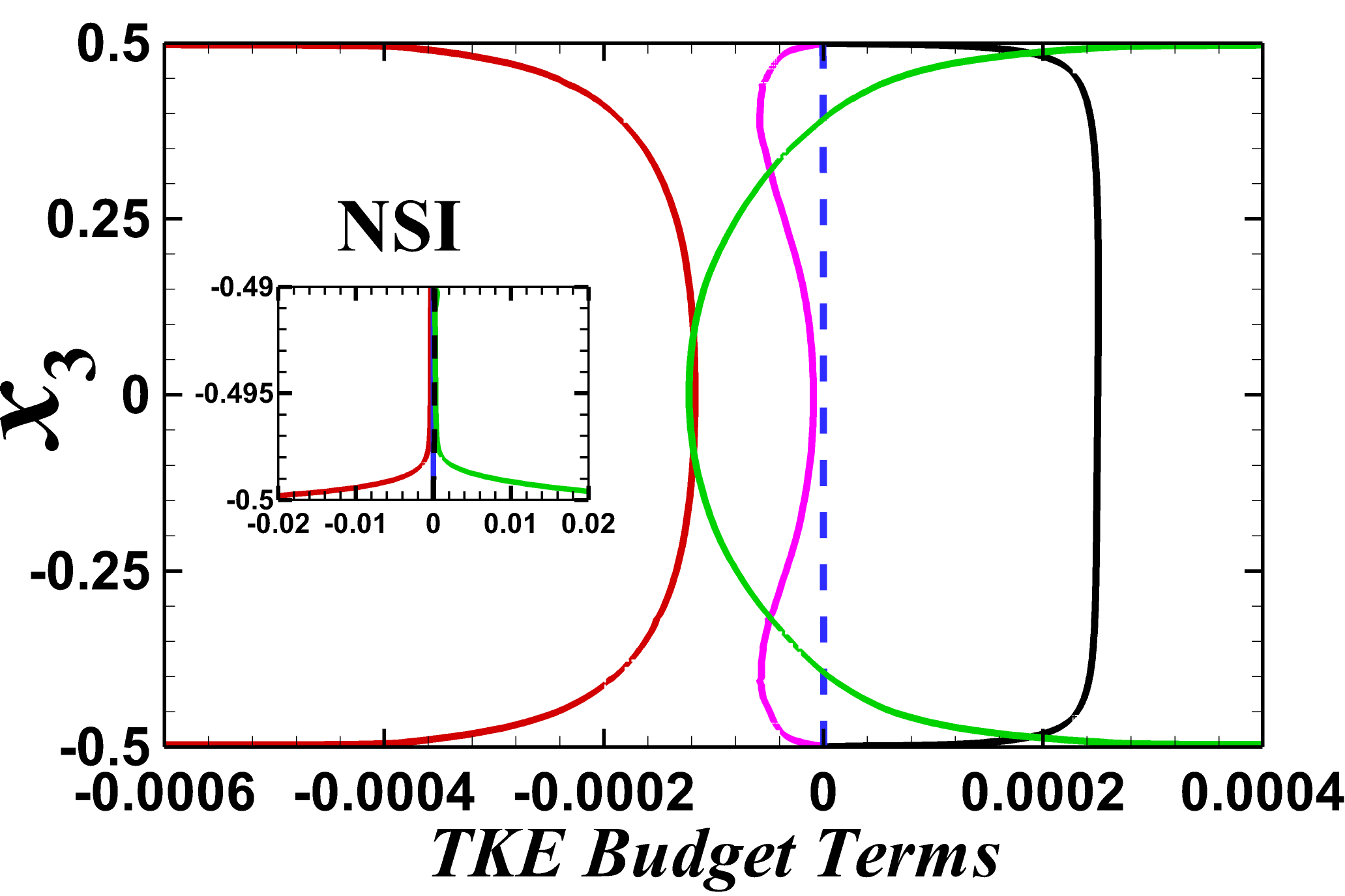}
(f) \includegraphics[width=0.467\linewidth,trim={0cm 0cm 0cm 0cm},clip]{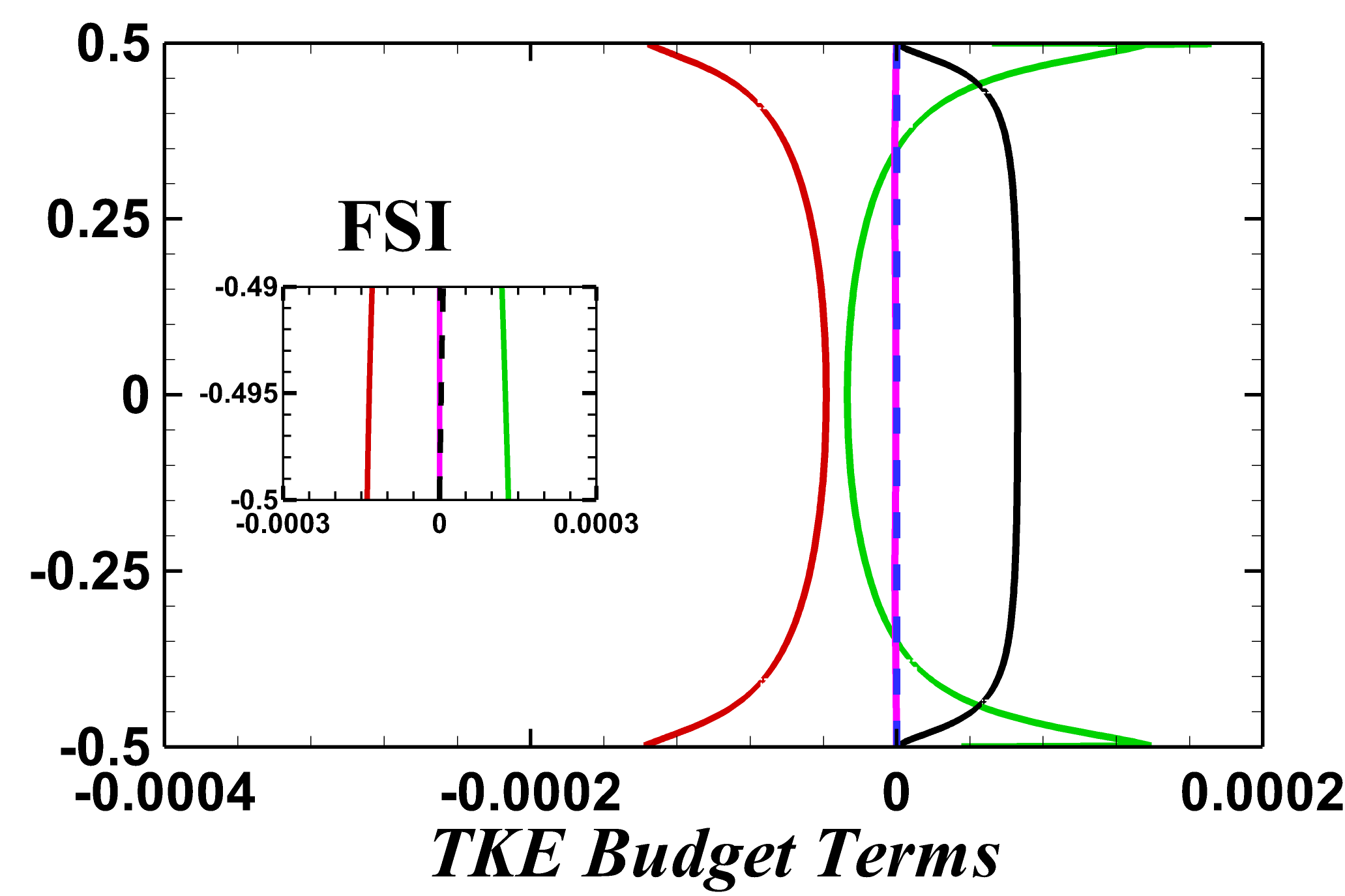}
\caption{Vertical variation of \textit{t.k.e} budget terms at $\mathcal{R}=3$ for both no-slip  (\textit{a}, \textit{c}, \textit{e}) and free-slip (\textit{b}, \textit{d}, \textit{f}) boundaries. Energy budget terms are presented for non-magnetic simulations (\textit{a}, \textit{b}) as well as dynamo simulations with both conducting (\textit{c}, \textit{d}) and insulating (\textit{e}, \textit{f}) wall. The budget terms averaged in the horizontal directions, as presented in equation \ref{eqn:TKE_budget}, are shown in the bulk and near the bottom plate(inset). The balance term signify the difference between left and right hand sides of equation \ref{eqn:TKE_budget}, and indicates sufficient accuracy of the present calculations. }
\label{fig:budget}
\end{figure*}

An analysis of the different energetic terms in the turbulent kinetic energy budget equation will illustrate the generation of \textit{t. k. e.} from thermal forcing, its dissipation and conversion to magnetic energy, elucidating the turbulent dynamo mechanism. We look into the significance of \textit{t.k.e.} budget terms and their dependence on boundary conditions, with particular attention to the near-wall features. In figure \ref{fig:budget}, the various terms in the \textit{t. k. e.} budget equation (see appendix \ref{app:Budget}) are plotted with the near-wall variation shown in the inset. All the terms are averaged in the horizontal planes, and their variation in the vertical direction is shown at $\mathcal{R}=3$ with varying boundary conditions. For NS and FS cases, in figure \ref{fig:budget}a and b, the primary balance is between the buoyant production ($\mathcal{B}$) and viscous dissipation($\epsilon_v$). The transport term $\partial T_j/\partial x_j$ acts to redistribute the energy and becomes zero when averaged over the volume. This term includes the combined effect of diffusive transport, pressure transport, and third-order correlation terms (see equation \ref{eqn:TKE_transport}). A detailed discussion of a typical energy budget in rotating convection with the individual treatment of these terms can be found in \citet{kunnen_2009}. In the absence of mean flow, the shear production term, $P$, is zero, whereas the unsteady term $\partial K/\partial t$ also becomes zero in the statistically stationary state. In the bulk, for the non-magnetic RC cases (NS and FS, shown in figures \ref{fig:budget}a and b), some part of the turbulent kinetic energy generated from buoyant production is converted to thermal energy by viscous dissipation, and the rest is redistributed by the transport term. Near the boundaries, this transport of kinetic energy transfers energy from the bulk towards the boundary layers, which undergoes viscous dissipation. Hence, the primary balance is between the dissipation and transport terms near the plates. Due to the presence of viscous Ekman layers in the NS case, the viscous dissipation (and therefore the kinetic energy transport) increases by two orders of magnitude near the plates compared to their bulk values as shown in the inset of figure \ref{fig:budget}a. Such an order of magnitude jump is not present for the FS case in figure \ref{fig:budget}b owing to the absence of the Ekman boundary layer. \\

For dynamos, the magnetic production of turbulent kinetic energy $\mathcal{P}$ appears to represent the work done by the Lorentz force on the velocity field to produce turbulent kinetic energy. A negative value of this term indicates transfer of energy from the velocity field to the magnetic field and vice-versa. This term is responsible for converting some part of the kinetic energy to magnetic energy to sustain dynamo action. For all the dynamo cases (NSC, NSI, FSC, and FSI), $\mathcal{P}$ appears as an additional sink of turbulent kinetic energy that balances the buoyant production of kinetic energy together with viscous dissipation. The behavior of magnetic production is similar to viscous dissipation for the NSC case. Here the budget terms show a peak near the edge of the Ekman layer as shown in the inset of \ref{fig:budget}c. For NSI boundary condition in figure \ref{fig:budget}e, the magnetic production term is not significant near the midplane but increases towards the boundary to reach a peak near $x=\pm0.4$ before it decreases to zero at the wall. For FSC and FSI conditions, the contribution from this term is not significant to the overall budget as plotted in figures \ref{fig:budget}d and f. This magnetic production term in the turbulent kinetic energy budget comprises of three components $\mathcal{P}_1$, $\mathcal{P}_2$ and $\mathcal{P}_3$ as expressed in equation \ref{eqn:PTKE}. Out of these, the first two magnetic production terms involving the mean magnetic energy and its gradients $\mathcal{P}_1$ and $\mathcal{P}_2$ are found to be small compared to $\mathcal{P}_3$ at $\mathcal{R}=3$. It should be noted here that for lower convective supercriticality, the mean magnetic field may become strong compared to the turbulent magnetic field, and therefore the terms $\mathcal{P}_1$ and $\mathcal{P}_2$ may significantly contribute to magnetic production. The transport of kinetic energy by the mean and fluctuating magnetic fields, as represented by the last two terms in the RHS of equation \ref{eqn:TKE_transport}, are also found to be small compared to the pressure transport and turbulent transport terms (figure not presented).\\

\begin{figure}%[tbhp]
\centering
(a)\includegraphics[width=0.7\linewidth,trim={0.2cm 8.15cm 0.2cm 6.15cm},clip]{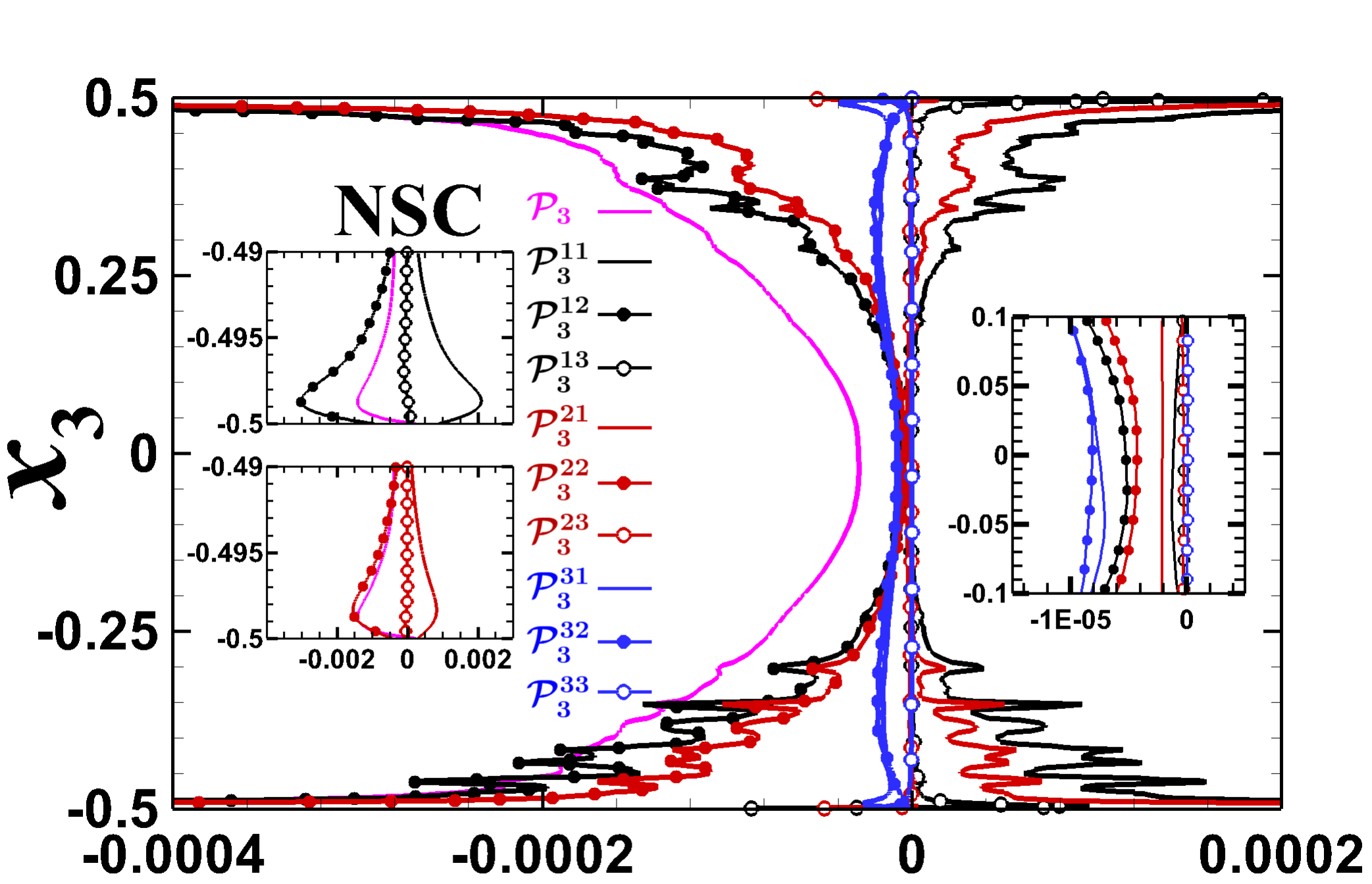}
(b)\includegraphics[width=0.7\linewidth,trim={0.2cm 6.15cm 0.2cm 6.15cm},clip]{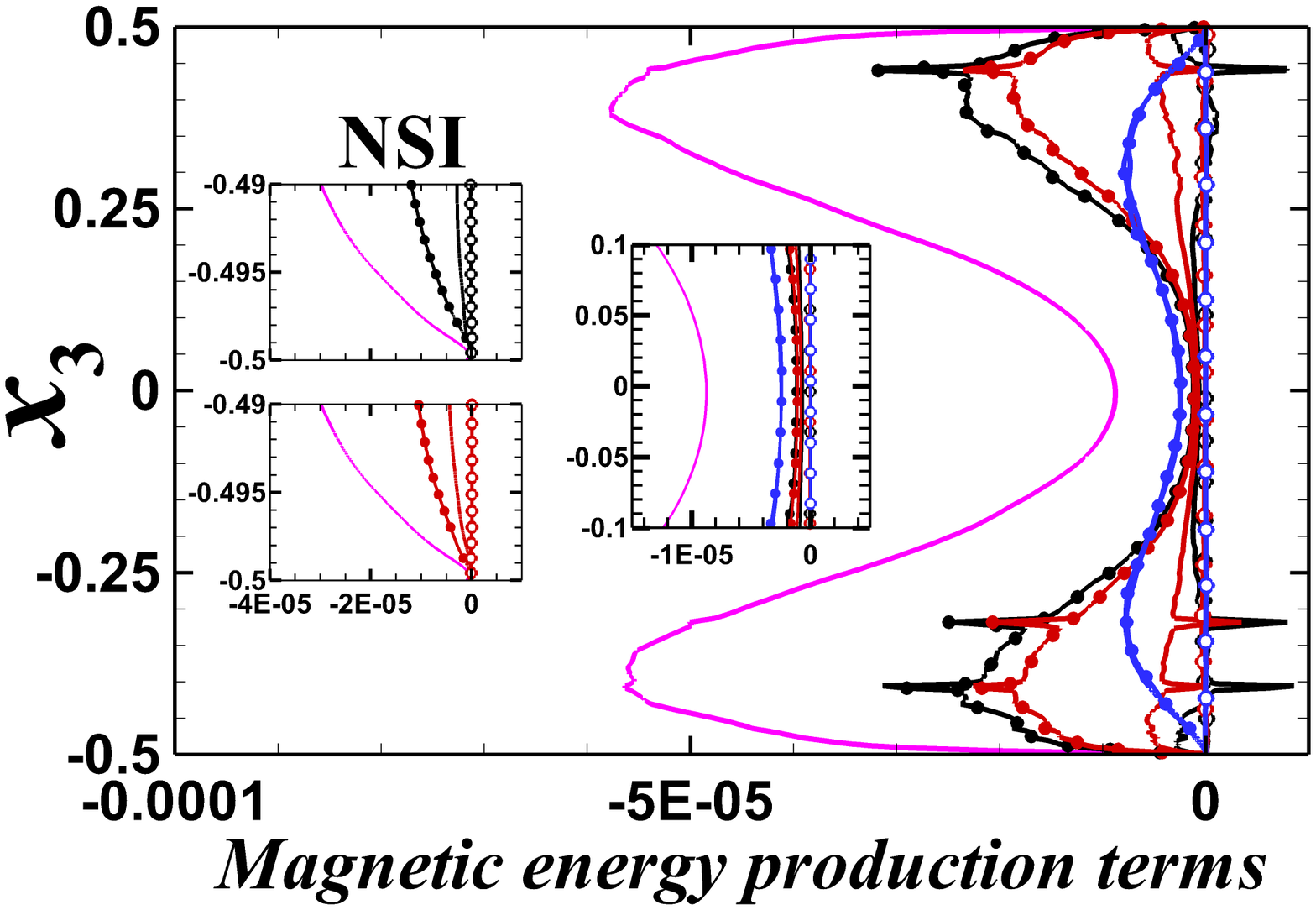}
\caption{Variation of magnetic production terms for (a) NSC and (b) NSI case at $\mathcal{R}=3$. The components of the work done by Lorentz force $\mathcal{P}_3$,  are represented by the colors black for $i=1$, red for $i=2$ and blue for $i=3$. Here line without symbols , lines with filled and open circles represent $j=1$, $j=2$ and $j=3$ respectively. }
\label{fig:pm}
\end{figure}

The dominant term among the magnetic production terms, $\mathcal{P}_3$, represents the work done by the fluctuating magnetic field on the fluctuating velocity field to produce turbulent kinetic energy. It is interesting to study this term for elucidating the mechanism of energy exchange between these two vector fields. The nine components of this term are represented by $\mathcal{P}^{ij}_3=\overline{B^{\prime}_{i}B^{\prime}_{j}\partial u^{\prime}_{i}/\partial x_{j}}$ in figure \ref{fig:pm} for NSC and NSI cases where magnetic production makes significant contribution to the turbulent kinetic energy budget. Among these nine components, the three terms involving the vertical gradient of the fluctuating velocity field, $\mathcal{P}^{i3}_3=\overline{B^{\prime}_{i}B^{\prime}_{3}\partial u^{\prime}_{i}/\partial x_{3}}$ where $i=1,2,3$, are small due to Taylor-Proudman constraint on the velocity field imposed by rotation that suppresses changes in the vertical direction. Interestingly, for the NSC case, there are components with both positive and negative values in figure \ref{fig:pm}a. Here, the components $\mathcal{P}^{11}_3$ and $\mathcal{P}^{21}_3$ become positive away from the midplane, indicating a transfer of energy from the magnetic field to the velocity field. Vertical variation of the terms $\mathcal{P}^{12}_3$ and $\mathcal{P}^{22}_3$ are similar but opposite, with negative values away from the mid-plane indicating transfer of energy from the velocity to the magnetic field. The last two negative terms, being larger than the positive terms, provide a bias in the direction of energy transfer so that the system can extract kinetic energy of the fluid and convert it to magnetic energy to sustain dynamo action. However, in the mid-plane, all the terms are negative. Near the wall, these four terms exhibit peaks near the edge of the Ekman boundary layer where the values are two orders of magnitude higher than mid-plane. The two remaining terms $\mathcal{P}^{31}_3$ and $\mathcal{P}^{32}_3$ are small compared to the dominant terms near the wall but contribute most to the energy transfer at the midplane. In the NSI case, all the components of work done by Lorentz force $\mathcal{P}^{ij}_3$ are negative, indicating unidirectional energy transfer from the velocity field to the magnetic field. This indicates that the imposed boundary conditions dictate the energy exchange mechanism between velocity and magnetic fields both in the bulk and near the boundaries. \\

The volume-averaged budget of turbulent magnetic energy, reduces to a balance between the magnetic energy production and Joule dissipation, $\langle\mathcal{P}\rangle=\langle\epsilon_j\rangle$, for statistically stationary turbulence. This signifies that the part of \textit{t. k. e.} converted to magnetic energy, ultimately converts to thermal energy via Joule dissipation. The Ohmic fraction, defined by the ratio of Joule dissipation to total dissipation $\langle\epsilon_j\rangle/\langle\epsilon\rangle$, is presented in tables \ref{tab:nsc}-\ref{tab:fsi}. However, the Ohmic fraction does not show any particular trend in the range $\mathcal{R}=2-20$. The global energy balance between buoyant production $\langle\mathcal{B}\rangle$ and total dissipation $\langle\epsilon\rangle$ is also presented in these tables, where the values have been scaled by a factor of $10^4$. The difference between the two terms never exceeds $5\%$ of $\langle\mathcal{B}\rangle$ indicating the accuracy of the present DNS in capturing all the energy containing scales. The vertical trends of the terms in \textit{t. k. e.} equation remains similar to that presented in figure \ref{fig:budget} over the range of thermal forcing studied here. The volume-averaged buoyancy flux and dissipation increases by an order of magnitude in the range $\mathcal{R}=2-20$ as seen in tables \ref{tab:nsc}-\ref{tab:fsi}.\\

\subsection{Heat transfer behavior}\label{sec:heattransfer}

\begin{figure}%[tbhp]
\centering
\includegraphics[width=0.75\linewidth,trim={0 6.15cm 0 6.15cm},clip]{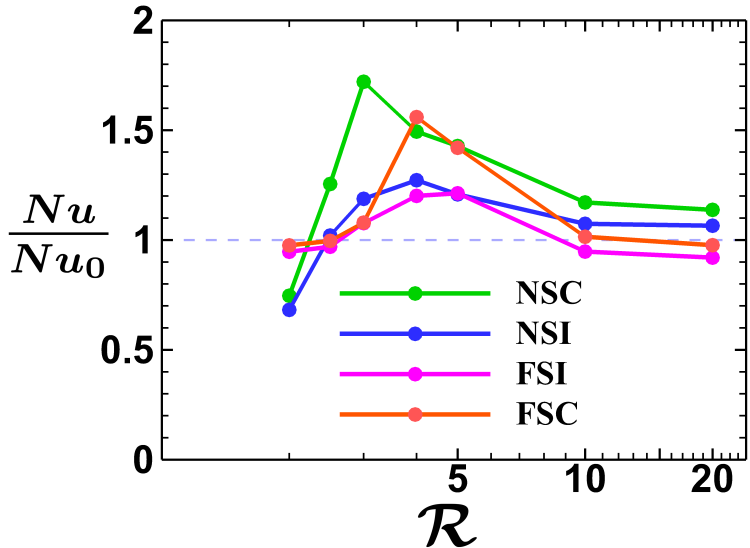}
\caption{Variation of Nusselt Number ratio $Nu/Nu_0$ as a function of convective supercriticality for different boundary conditions. Here $Nu_0$ represents the Nusselt number in non-magnetic simulations.}
\label{fig:nusselt}
\end{figure}

Finally, we look into the heat transfer behavior of the dynamo simulations with increasing thermal forcing under the influence of different boundary conditions. In figure \ref{fig:nusselt}, the Nusselt number ratio, $Nu/Nu_0$, signifies the change in heat transport due to dynamo action compared to the non-magnetic simulation. For NSC, a peak is found near $\mathcal{R}=3$ with more than $72\%$ enhancement in heat transfer. A peak in Nusselt number ratio is also found for the NSI case at $\mathcal{R}=4$, with a $27\%$ increase in heat transport compared to non-magnetic rotating convection. For FSC and FSI cases, the heat transfer ratio also peaks at $\mathcal{R}=4$ and $\mathcal{R}=5$ with enhancement up to $44\%$ and $21\%$ respectively. The ratio of viscous dissipation in magnetic and no-magnetic simulation $\langle\epsilon_v\rangle/\langle\epsilon_0\rangle$ is also presented in the table, which shows a similar trend with the Nusselt number ratio. For conducting boundaries, the peaks in heat transfer ratio and the viscous dissipation ratio occurs at the same value of $\mathcal{R}$. We note here that the overall energy balance of the system leads to an exact relation between the Nusselt number and the total dissipation, $Nu=1+\sqrt{RaPr}\langle\epsilon\rangle$. Hence the dissipation ratio should behave similarly as the Nusselt number ratio with varying $\mathcal{R}$ as seen in tables \ref{tab:nsc}-\ref{tab:fsi}. A local magnetorelaxation of rotational constraint, due to enhanced Lorentz force in the thermal boundary layer (see figure \ref{fig:forces}c), has been shown to be the reason for the rise in heat transfer in the NSC case\citep{naskar_2021}. However, the build-up of the Lorentz force required for this magnetorelaxation process is not present for any other combinations of boundary conditions. Notably, in the NSI case, the Lorentz force in the bulk is higher compared to other cases. Indeed, the volume-averaged Elsasser number, $\Lambda_V$ in table \ref{tab:nsi} shows a peak at $\mathcal{R}=4$ that correlates well with the heat transfer behavior. Therefore, the global relaxation of the Taylor-Proudman constraint is a possible mechanism for heat transfer enhancement.  For the free-slip cases, the heat transfer enhancement in dynamos is achieved by suppression of LSVs by the small-scale magnetic field, as discussed in the next paragraph. It is interesting to note that the peak in heat transfer ratio appears within a narrow range of convective supercriticality $\mathcal{R}=3-5$ depending on the conditions at the boundary. Further investigations will be necessary to elucidate the appearance of the peak in heat transfer ratio for different boundary conditions. However, the heat transport properties of the dynamo depend on the boundary conditions and the associated structure of the magnetic field and dynamical balances both in the bulk and in the boundary layer.\\

\begin{figure}%[tbhp]
\centering
(a)\includegraphics[width=0.485\linewidth,trim={0.2cm 5.2cm 1cm 5cm},clip]{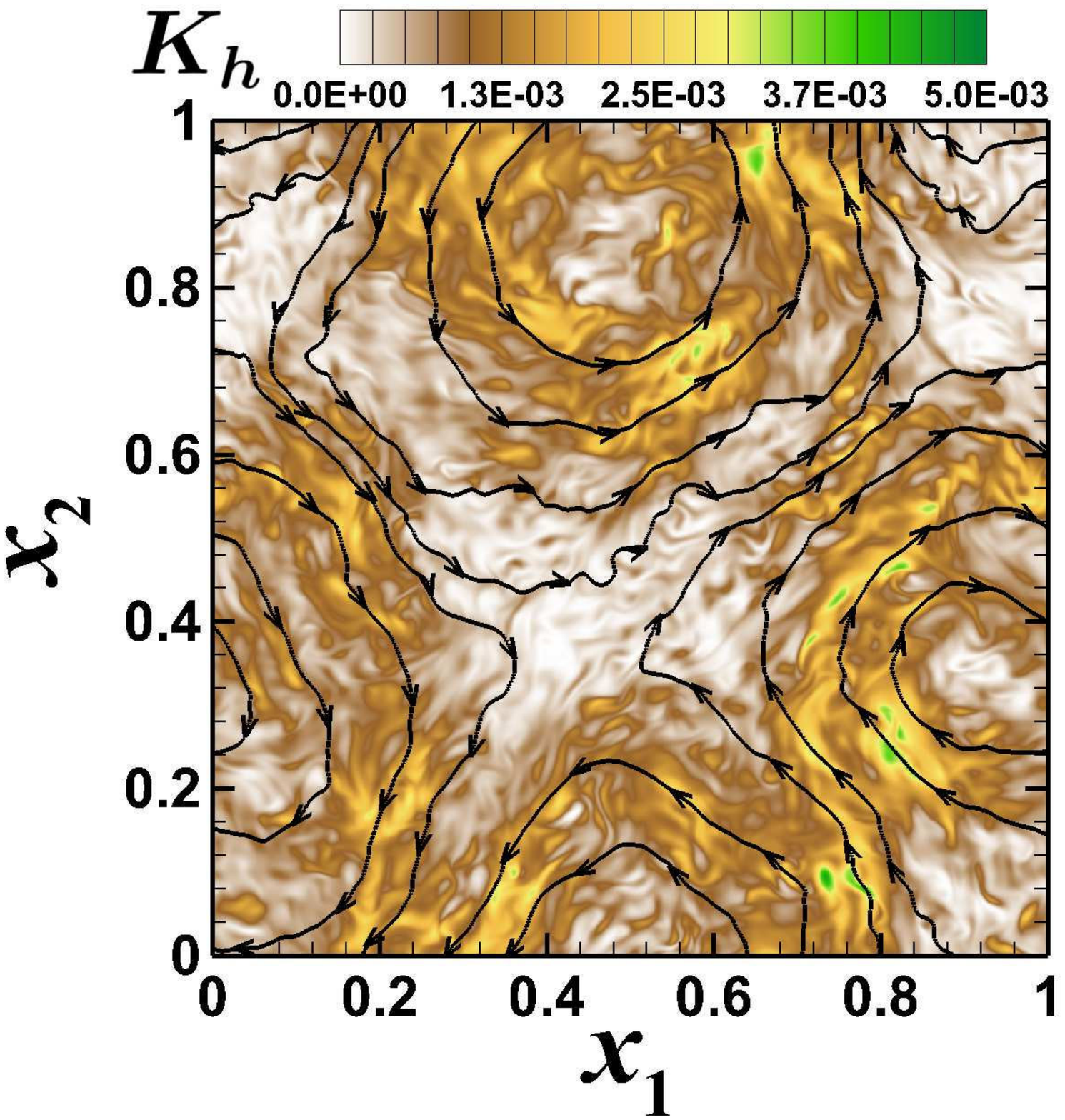}
(b)\includegraphics[width=0.445\linewidth,trim={2cm 5.2cm 1cm 5cm},clip]{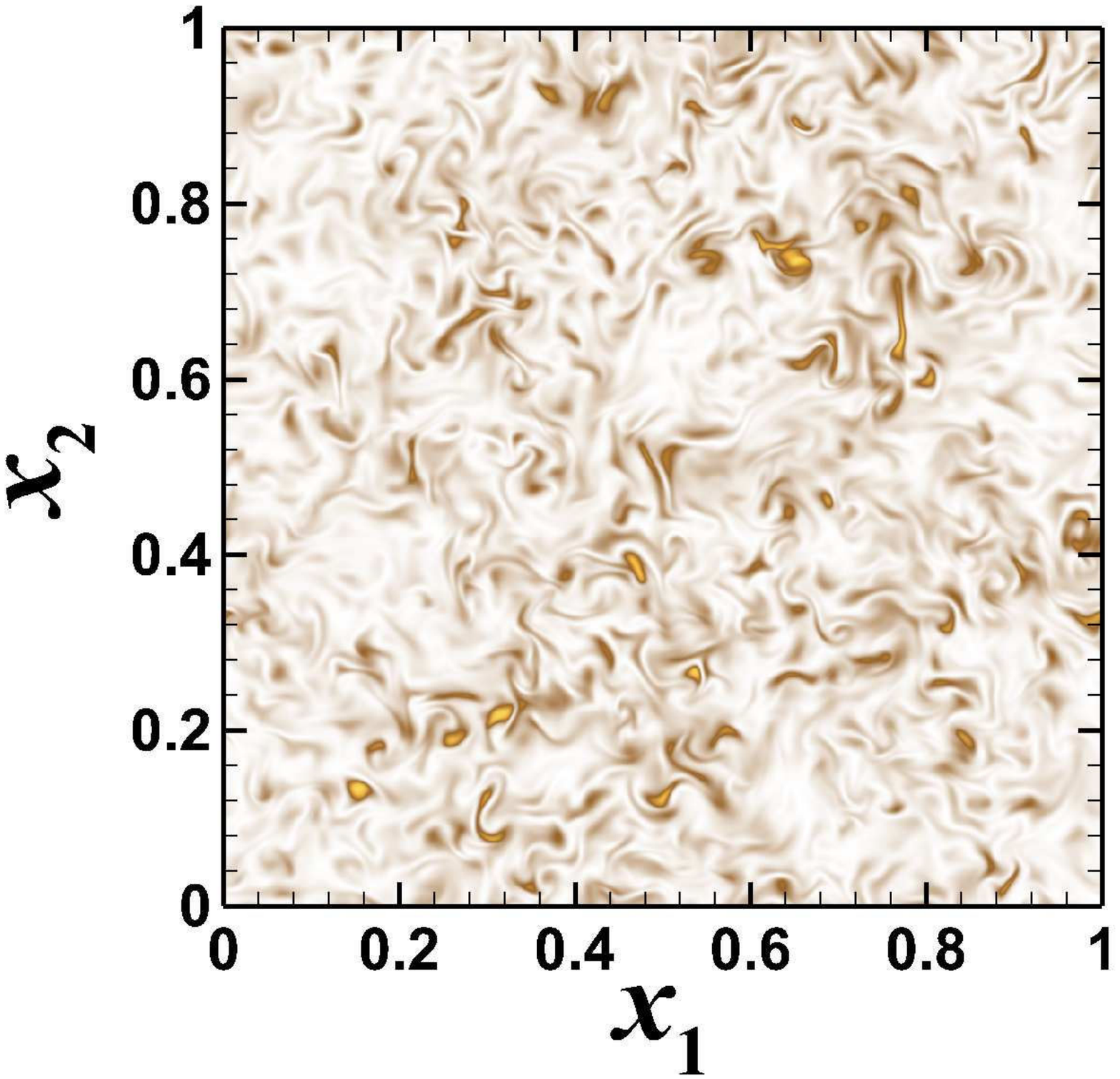}\\
(c)\includegraphics[width=0.48\linewidth,trim={0.2cm 3cm 1cm 5cm},clip]{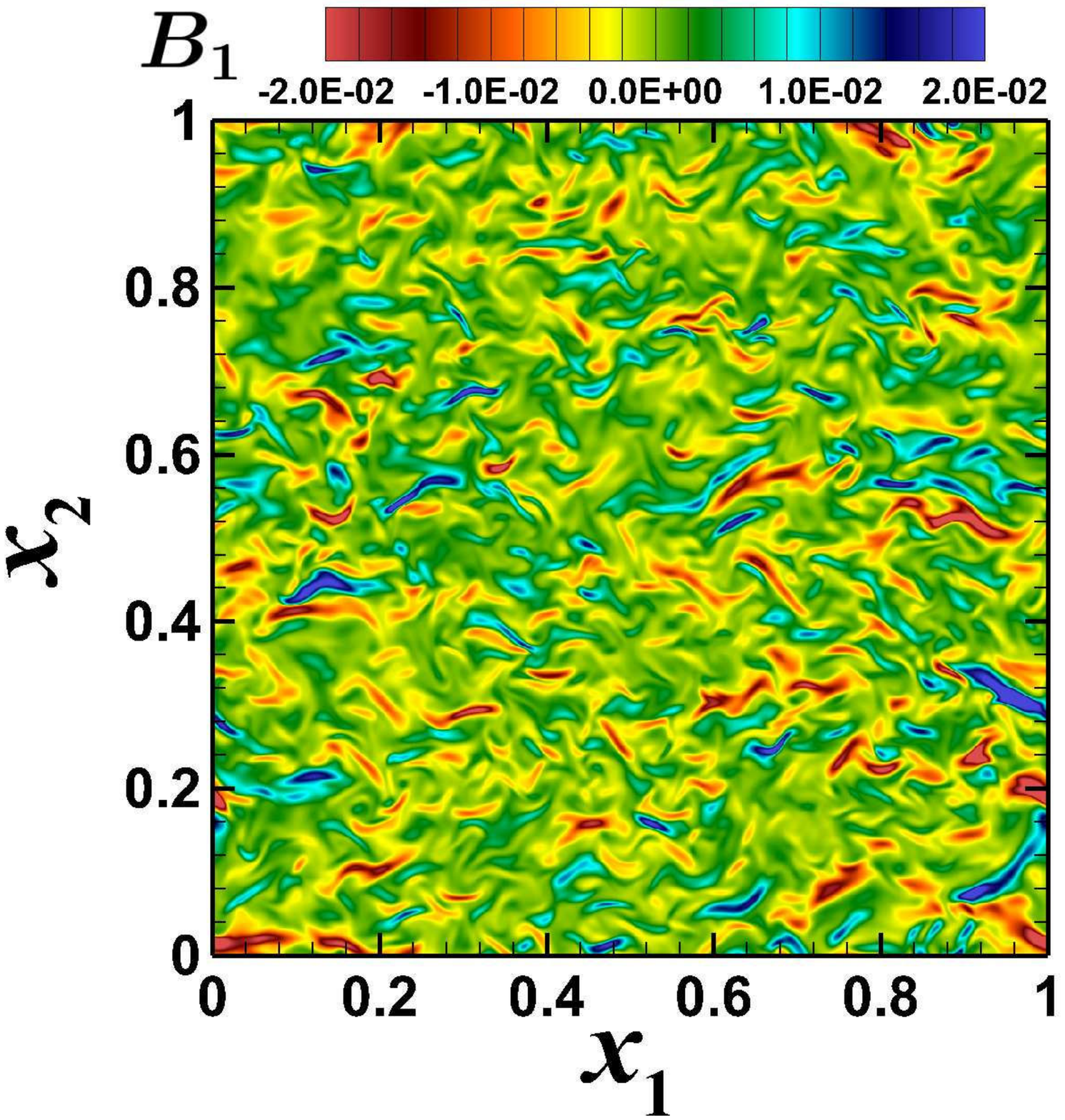}
(d)\includegraphics[width=0.45\linewidth,trim={0.5cm 3cm 0.2cm 5cm},clip]{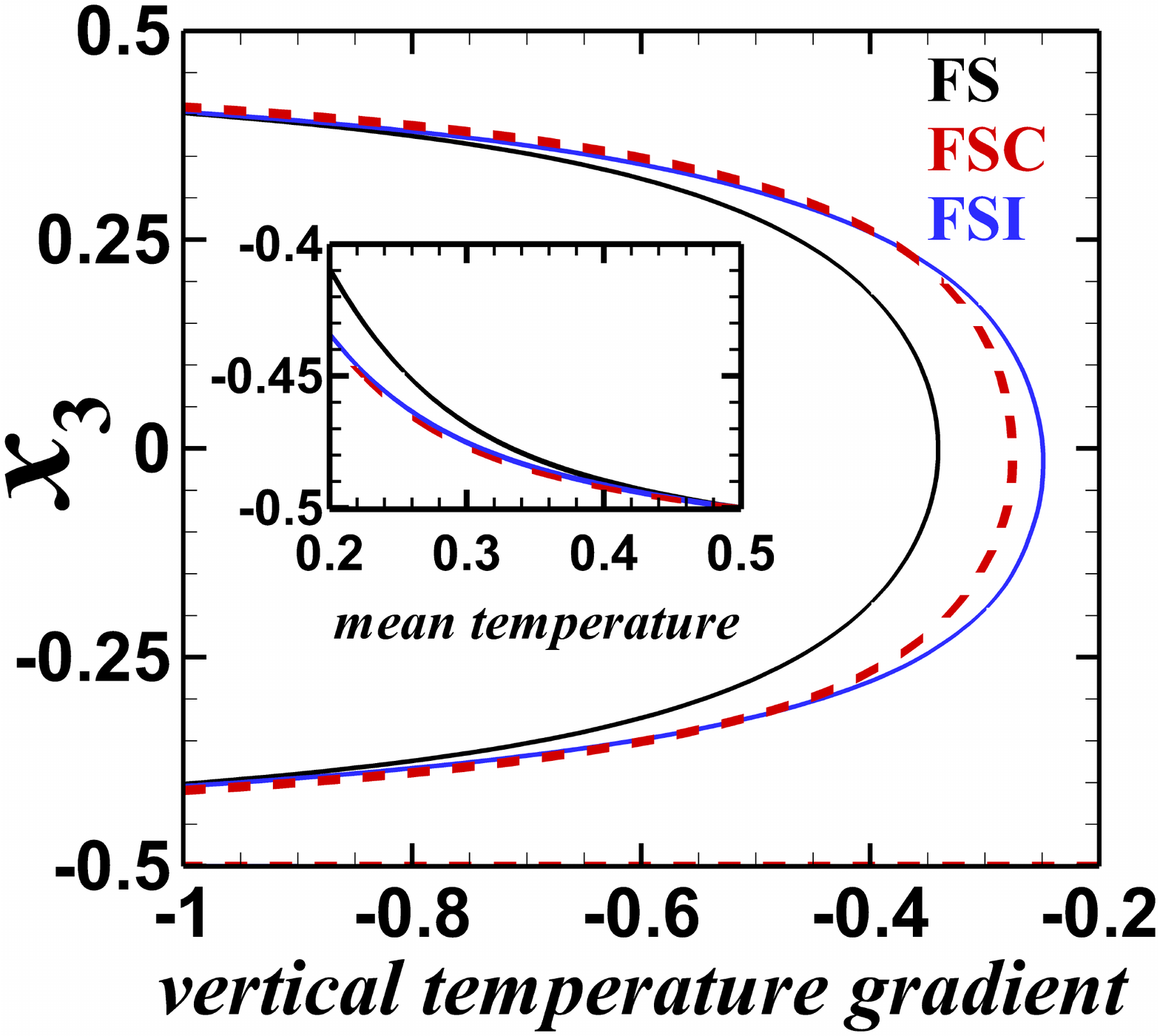}
\caption{Contours of horizontal turbulent kinetic energy, $K_h=1/2(u_{1,rms}^2+u_{2,rms}^2)$  for (a) FS and (b) FSC cases at $\mathcal{R}=4$. The instantaneous plots are shown in the horizontal mid-plane at $x_3=0.5$ to illustrate the presence of LSVs in (a) FS which is superimposed with streamlines. The dynamo simulation with FSC conditions generates small scale velocity and magnetic fields as illustrated by contours of \textit{t.k.e} and magnetic field $B_1$ in (b) and (c) respectively. The vertical gradient of time and horizontally averaged mean temperature profile is shown in (d). Here, the near wall temperature profiles are shown in the inset.}
\label{fig:lsv}
\end{figure}

The heat transfer enhancement in dynamos with free-slip boundary conditions with respect to non-magnetic rotating convection can be attributed to the absence of LSVs in the former. In the FS cases, we find the presence of depth-independent, long-lived (compared to the free-fall time scale) LSVs for $\mathcal{R}\geq4$, corroborating the findings of \citet{guervilly_2014}. Similar to freely decaying 2D turbulence, the formation of LSVs in rotating convection involves an upscale energy transfer from the convective eddies to large-scale barotropic modes \citep{favier_2014,rubio_2014}. The flow associated with these structures are nearly 2D, that can be visualized from the contours of horizontal turbulent kinetic energy, $K_h=1/2(u_{1,rms}^2+u_{2,rms}^2)$, which tends to be high in the shear layer around the core of these vortices as shown in figure \ref{fig:lsv}a for $\mathcal{R}=4$. An animation of the LSV, visualized by the time evolution of the $K_h$ contours, can be found in the supplementary video. Here, we find a pair of cyclonic and anti-cyclonic vortices centered around $(0.48,0.87)$ and $(0.93,0.34)$ respectively as depicted in the horizontal mid-plane at $x_3=0.5$. The degree to which the two-dimensionality of the flow has been induced by the presence of an LSV can be measured by the ratio $\tau=(u_{1,rms}^2+u_{2,rms}^2+u_{3,rms}^2)/3u_{3,rms}^2$, as reported in table \ref{tab:nsc}-\ref{tab:fsi}. This kinetic energy ratio for non-magnetic simulations $\tau_{0}$ is also reported in \ref{tab:nsi} and \ref{tab:fsc} for NS and FS cases respectively. As the kinetic energy associated with the horizontal flow in the LSV is higher than the kinetic energy in the vertical flow, the value of $\tau$ increase to values much higher than unity in the FS cases for $\mathcal{R}\geq4$. We have also found LSVs with no-slip boundary conditions without magnetic field (NS case) for $\mathcal{R}=10\  \text{and}\  20$, which supports the findings of \citet{guzman_2020}, that LSVs appear in rotating convection irrespective of the kinematic boundary condition. However, the presence of no-slip boundaries suppresses the formation of LSVs until higher convective supercriticality ($\mathcal{R}\geq10$) compared to the free-slip cases ($\mathcal{R}\geq4$). \\

\citet{guervilly_2017} reported that the formation of LSVs can be suppressed by the presence of a small-scale magnetic field for $Re_m\geq550$, by disrupting the correlations between convective vortices that leads to upscale transfer of energy. This should be the reason that we do not find any LSVs in our dynamo simulations with free-slip conditions that operate for $Re_m\geq641$. The combined effect of no-slip boundary condition and the presence of magnetic field has suppressed LSVs in the NSC and NSI cases as well, as reflected by the value of $\tau$ in table \ref{tab:nsc} and \ref{tab:nsi}. \citet{maffei_2019} used an asymptotic magnetohydrodynamic model to study the effect of an imposed magnetic field on the inverse cascade of energy in rapidly rotating convection in the turbulent geostrophic regime. The presence of LSVs in this regime was quantified by an interaction parameter $N$, which signifies the relative strengths of the Lorentz force and the non-linear advection. For $N\geq0.013$ the magnetic field disrupts the upscale energy transfer, and hence LSVs can only be present below this limit. We have calculated the interaction parameter as a ratio of volume-averaged \textit{r.m.s} magnitudes of the Lorentz force and the non-linear advection terms, and found the ratio to be of $O(0.1)$ for all our dynamo simulations. Though the self-excited magnetic fields in our dynamo simulations are not directly comparable to the externally imposed field used by \citet{maffei_2019}, the magnetic quenching of the inverse cascade is a plausible mechanism for the disappearance of LSVs in our dynamo simulations. The horizontal turbulent kinetic energy and the $x_1$-component of the magnetic field for FSC cases depicted in figure \ref{fig:lsv}b and c shows small-scale fields without any trace of LSVs. The presence of LSVs can disrupt the vertical mixing by transferring energy to horizontal barotropic modes \citep{guervilly_2014}. In figure \ref{fig:lsv}d, this phenomenon results in an increased gradient of temperature in the bulk for the FS case with LSV, compared to the FSC and FSI cases without LSV. The temperature profile near the wall (see inset) shows higher thermal boundary layer thickness in the FS case compared to the dynamo cases, indicating a decrease in heat transfer efficiency near the wall in the FS case. The thermal plumes near the wall get swept away by the horizontal flow associated with LSVs, which can interrupt the vertical transport of heat by the plumes, leading to a decrease in heat transfer efficiency.\\          

\section{Conclusions}\label{sec:conclusion}

We have performed high fidelity direct numerical simulations of Rayleigh-B\'enard convection-driven dynamos under four combinations of kinematic and magnetic boundary conditions in the rotation dominated regime $\mathcal{R}=2-20$ at a fixed rotation rate $E=5\times10^{-7}$. The dynamo simulations with four combinations of no-slip and free-slip kinematic boundary conditions along with perfectly conducting and insulating magnetic boundary conditions (NSC, NSI, FSC, and FSI) are compared against non-magnetic simulations (NS and FS) to study the behavior of the magnetic field generated in each case. Our previous study \citep{naskar_2021} reported the existence of optimal heat transfer enhancement with respect to non-magnetic convection due to dynamo action for NSC boundary conditions at $\mathcal{R}=3$. Therefore, we have chosen this case to compare the dynamical balance, energy budget, and heat transfer behavior of the dynamos for different boundary conditions. \\

Apart from the magnetic field structure, the flow and thermal field of the dynamos are found to depend on the boundary conditions. Ekman pumping significantly enhances the velocity and thermal fluctuations in the dynamos, increasing the \textit{r.m.s} velocities and temperatures both in the bulk and near the boundary layers. It also leads to a three order of magnitude increase in enstrophy, signifying increased strength of the vortical plumes inside the Ekman layer with no-slip boundary condition. The relative helicity also exhibits a peak near the Ekman layer, illustrating strong correlations between the velocity and vorticity fields. The perfectly conducting boundary conditions can trap the magnetic field near the boundaries to prevent its escape and make it purely horizontal. This leads to a jump of \textit{r.m.s} horizontal field strength near the boundaries for the NSC case. The \textit{r.m.s} field strength remains higher than the mean-field for all combinations of boundary conditions in the investigated range of thermal forcing.\\

The leading order force balance in the dynamos remains geostrophic, similar to our non-magnetic rapidly rotating convection simulations, irrespective of the boundary conditions. At $\mathcal{R}=3$, the quasi-geostrophic balance corresponds to a CIA balance between the ageostrophic part of Coriolis force, inertia, and buoyancy forces in the non-magnetic simulations. The extent to which the Lorentz force modulates the quasi-geostrophic balance is decided by the boundary conditions in the dynamo simulations. For NSC conditions, a build-up of Lorentz force is seen near the walls, whereas this force dominate the quasi-geostrophic balance in the bulk for NSI conditions. For free-slip conditions, Lorentz force is much weaker compared to the no-slip dynamos. \\

The budget of turbulent kinetic energy for the non-magnetic cases exhibits an overall balance between the buoyant production and viscous dissipation, where the transport terms act to redistribute the energy in the system. For the NS case, the viscous dissipation increases by two orders of magnitude from its bulk value due to viscous action inside the Ekman layer. For dynamo simulations, part of the turbulent kinetic energy is converted to magnetic energy via work done by the Lorentz force, which is ultimately converted to thermal energy by Joule dissipation. This additional term, signifying the production of magnetic energy in the dynamo simulations, is considerably weaker with free-slip boundaries than no-slip boundaries. A break up of components that constitutes this production of magnetic energy term (which can also be interpreted as the work done by Lorentz force) reveals that, the energy flows both ways, from velocity field to the magnetic field and vice-versa, when NSC conditions are imposed at the boundary. Conversely, the flow of energy is unidirectional, from kinetic energy to magnetic energy, when NSI boundary condition is used, indicating the decisive role of the boundary conditions on the mechanism of energy transfer.\\

Another interesting finding of our study is the enhancement of heat transfer in the dynamo simulations with respect to non-magnetic rotating convection. The heat transfer enhancement reaches a peak in the range $\mathcal{R}=3-5$ for all the combinations of boundary conditions. We find that the magnetorelaxation of the rotational constraint by the Lorentz force is the mechanism for heat transfer enhancement for the no-slip cases, whereas, for the free-slip dynamos, suppression of LSVs by magnetic field leads to the increased efficiency of heat transport. We have found LSVs in our non-magnetic simulations that are known to deteriorate the heat transfer, as reported in literature \citep{guervilly_2014}. By comparing our results with existing literature, we conclude that the magnetic quenching of LSVs by the magnetic field is the possible reason for heat transfer enhancement in our free-slip dynamo simulations. An interesting extension of the present study will be the search for power-law scaling of the heat transfer and flow speed with the thermal forcing in such convection-driven dynamos and the effect of boundary conditions on the power-law exponent and the prefactor.  \\

\backsection[Supplementary data]{\label{SupMov}A Supplementary movie of Large scale vortex is available at \\https://doi.org/**.****/jfm.***...}

\backsection[Acknowledgements]{We gratefully acknowledge the support of the Science and Engineering Research Board, Government of India grant no. SERB/ME/2020318. We also want to thank the Office of Research and Development, Indian Institute of Technology Kanpur for the financial support through grant no. IITK/ME/2019194. The support and the resources provided by PARAM Sanganak under the National Supercomputing Mission, Government of India at the Indian Institute of Technology, Kanpur are gratefully acknowledged.}

%\backsection[Funding]{This research received no specific grant from any funding agency, commercial or not-for-profit sectors.}

\backsection[Declaration of interests]{ The authors report no conflict of interest.}

%\backsection[Data availability statement]{The data that support the findings of this study are openly available in [repository name] at http://doi.org/[doi], reference number [reference number].}

%\backsection[Author ORCID]{Souvik Naskar, https://orcid.org/0000-0003-0445-8417; Anikesh Pal, https://orcid.org/****-****-****-****}

\backsection[Author contributions]{  The authors contributed equally to analysing data and reaching conclusions, and in writing the paper.}

\appendix

\section{}\label{app:Budget}

The turbulent kinetic energy equation can be derived including the terms involving work done by the Lorentz forces. We note here that the Coriolis force is a pseudo force and does not enter the balance directly. The evolution of the horizontally averaged \textit{t.k.e} (see section \ref{sec:stat} for the definition of the averages),  can be written as the following.

\begin{equation}\label{eqn:TKE_budget}
    \frac{dK}{dt}=-P+\mathcal{B}-\epsilon_v-\frac{\partial\mathcal{T}_{j}}{\partial x_j}+\mathcal{P};
\end{equation}

where 

\begin{equation}\label{eqn:TKE_terms}
\bf \begin{split}
     K=\frac{1}{2}\overline{u^{\prime}_i{}u^{\prime}_i{}};\qquad
     P=-\overline{u^{\prime}_iu^{\prime}_j}\frac{\partial\Bar{u_i}}{\partial x_j};\qquad 
     \mathcal{B}=\overline{u^{\prime}_3\theta^{\prime}};\qquad
     \epsilon_v=\sqrt{\frac{Pr}{Ra}}\overline{\frac{\partial u^{\prime}_i}{\partial x_j}\frac{\partial u^{\prime}_i}{\partial x_j}};
 \end{split}
\end{equation}

are the turbulent kinetic energy, turbulent production, buoyancy flux and the viscous dissipation. The transport of $K$ is given below

 \begin{equation}\label{eqn:TKE_transport}
    \mathcal{T}_j=\overline{u^{\prime}_jp^{\prime}}+\frac{1}{2}\overline{u^{\prime}_iu^{\prime}_iu^{\prime}_j}-\sqrt{\frac{Pr}{Ra}}\frac{\partial ^2K}{\partial x_j \partial x_j}-\Bar{B_j}\overline{u^{\prime}_i{}B^{\prime}_i}-\overline{u^{\prime}_iB^{\prime}_iB^{\prime}_j}
\end{equation}

The shear production term $P$ in equation \ref{eqn:TKE_terms} is negligible in the absence of a mean flow in the present simulations. However, shear production may still arise in the presence of a mean vertical motion through the turbulent transport term in equation \ref{eqn:TKE_transport} \citep{Kerr_2001}.The last term in the RHS of equation \ref{eqn:TKE_budget} is the production of $K$ due to work done by Lorentz force.

 \begin{equation}\label{eqn:PTKE}
 %\begin{split}
     \mathcal{P}_{1}=-\Bar{B_j}\overline{B^{\prime}_i\frac{\partial u^{\prime}_i}{\partial x_j}};\quad
     \mathcal{P}_{2}=\overline{u^{\prime}_iB^{\prime}_j}\frac{\partial \Bar{B_i}}{\partial x_j};\quad
     \mathcal{P}_{3}=-\overline{B^{\prime}_iB^{\prime}_j\frac{\partial u^{\prime}_i}{\partial x_j}};\quad
     \mathcal{P}=\mathcal{P}_{1}+\mathcal{P}_{2}+\mathcal{P}_{3}
 %\end{split}
\end{equation}

 These magnetic production terms $\mathcal{P}_{1}\ \textrm{to}\ \mathcal{P}_{3}$ in equation \ref{eqn:PTKE} represent the exchange of energy between the velocity and the magnetic fields. For example, $\mathcal{P}_1$ signify the production of turbulent kinetic energy due to work done by the mean magnetic field on the fluctuating strain rate of the velocity field. Furthermore, $\mathcal{P}_2$ represents the production of turbulent kinetic energy due to mean magnetic field gradient, analogous to the shear production term $P$ in \ref{eqn:TKE_terms}. The amplification (or attenuation) of the magnetic energy, due to the work done by stretching (or squeezing) of fluctuating magnetic field lines by the fluctuating velocity gradients, is represented by the term $\mathcal{P}_3$. The terms $\mathcal{P}_{1}$ and $\mathcal{P}_{2}$, apart from the last two terms in equation \ref{eqn:TKE_transport}, representing the transport of kinetic energy by the magnetic field, remains small in the investigated range of parameters. \\

\bibliographystyle{jfm}
\bibliography{jfm}

\end{document}